\documentclass[preprint,12pt]{elsarticle}

\usepackage{amssymb}

\journal{XX}

\begin{document}

\begin{frontmatter}



\title{Explicit representation for a class of Type 2 constacyclic codes over the ring $\mathbb{F}_{2^m}[u]/\langle u^{2\lambda}\rangle$
with even length}


\author{Yuan Cao$^{a, \ b, \ c}$, Yonglin Cao$^{a, \ \ast}$, Hai Q. Dinh$^{d, \ e}$, Songsak Sriboonchitta$^{f}$,
Guidong Wang$^{a}$}

\address{$^{a}$School of Mathematics and Statistics,
Shandong University of Technology, Zibo, Shandong 255091, China

\vskip 1mm $^{b}$Hubei Key Laboratory of Applied Mathematics, Faculty of Mathematics and Statistics, Hubei University, Wuhan 430062, China

\vskip 1mm $^{c}$Hunan Provincial Key Laboratory of Mathematical Modeling and Analysis in Engineering, Changsha University of Science and Technology, Changsha, Hunan 410114, China

\vskip 1mm $^{d}$Division of Computational Mathematics and Engineering, Institute for Computational
       Science, Ton Duc Thang University, Ho Chi Minh City, Vietnam

\vskip 1mm $^{e}$Faculty of Mathematics and Statistics, Ton Duc Thang University, Ho Chi Minh City,
      Vietnam

\vskip 1mm $^{f}$Faculty of Economics, Chiang Mai University, Chiang Mai 52000, Thailand}
\cortext[cor1]{corresponding author.  \\
E-mail addresses: yuancao@sdut.edu.cn (Yuan Cao), \ ylcao@sdut.edu.cn (Yonglin Cao),
\ dinhquanghai@tdtu.edu.vn (H. Q. Dinh), \ songsakecon@gmail.com (S. Sriboonchitta), \ hbuwgd@163.com (G. Wang).}

\begin{abstract}
Let $\mathbb{F}_{2^m}$ be a finite field of cardinality $2^m$,
$\lambda$ and $k$ be integers satisfying $\lambda,k\geq 2$
and denote $R=\mathbb{F}_{2^m}[u]/\langle u^{2\lambda}\rangle$. Let $\delta,\alpha\in \mathbb{F}_{2^m}^{\times}$.
For any odd positive integer $n$, we give an explicit representation and enumeration for all distinct $(\delta+\alpha u^2)$-constacyclic codes over $R$ of length $2^kn$, and provide a clear formula to count the number of all these codes.
As a corollary, we conclude that
every $(\delta+\alpha u^2)$-constacyclic code over $R$ of length $2^kn$ is an ideal
generated by at most $2$ polynomials in the residue class ring $R[x]/\langle x^{2^kn}-(\delta+\alpha u^2)\rangle$.
\end{abstract}


\end{frontmatter}


\section{Introduction}
\noindent
 Algebraic coding theory deals with the design of error-correcting and error-detecting codes for the reliable transmission
of information across noisy channel. The class of constacyclic codes plays a very significant role in
the theory of error-correcting codes.  It includes as a subclass of the important class of cyclic codes, which has been well studied since the late 1950's. Constacyclic codes also have practical applications as they can be efficiently encoded with simple shift registers. This family of codes is thus interesting for both theoretical and practical reasons.

\par
  Let $\Gamma$ be a commutative finite ring with identity $1\neq 0$, and $\Gamma^{\times}$ be the multiplicative group of invertible elements of
$\Gamma$. For any $a\in
\Gamma$, we denote by $\langle a\rangle_\Gamma$, or $\langle a\rangle$ for
simplicity, the ideal of $\Gamma$ generated by $a$, i.e. $\langle
a\rangle_\Gamma=a\Gamma=\{ab\mid b\in \Gamma\}$. For any ideal $I$ of $\Gamma$, we will identify the
element $a+I$ of the residue class ring $\Gamma/I$ with $a$ (mod $I$) for
any $a\in \Gamma$ in this paper.

\par
   A \textit{code} over $\Gamma$ of length $N$ is a nonempty subset ${\cal C}$ of $\Gamma^N=\{(a_0,a_1,\ldots$, $a_{N-1})\mid a_j\in\Gamma, \
j=0,1,\ldots,N-1\}$. The code ${\cal C}$
is said to be \textit{linear} if ${\cal C}$ is an $\Gamma$-submodule of $\Gamma^N$. All codes in this paper are assumed to be linear. The ambient space $\Gamma^N$ is equipped with the usual Euclidian inner product, i.e.
$[a,b]=\sum_{j=0}^{N-1}a_jb_j$, where $a=(a_0,a_1,\ldots,a_{N-1}), b=(b_0,b_1,\ldots,b_{N-1})\in \Gamma^N$,
and the \textit{dual code} is defined by ${\cal C}^{\bot}=\{a\in \Gamma^N\mid [a,b]=0, \forall b\in {\cal C}\}$.
If ${\cal C}^{\bot}={\cal C}$, ${\cal C}$ is called a \textit{self-dual code} over $\Gamma$.
   Let $\gamma\in \Gamma^{\times}$.
Then a linear code
${\cal C}$ over $\Gamma$ of length $N$ is
called a $\gamma$-\textit{constacyclic code}
if $(\gamma c_{N-1},c_0,c_1,\ldots,c_{N-2})\in {\cal C}$ for all
$(c_0,c_1,\ldots,c_{N-1})\in{\cal C}$. Particularly, ${\cal C}$ is
called a \textit{negacyclic code} if $\gamma=-1$, and ${\cal C}$ is
called a  \textit{cyclic code} if $\gamma=1$.

\par
  For any $a=(a_0,a_1,\ldots,a_{N-1})\in \Gamma^N$, let
$a(x)=a_0+a_1x+\ldots+a_{N-1}x^{N-1}\in \Gamma[x]/\langle x^N-\gamma\rangle$. We will identify $a$ with $a(x)$ in
this paper. Then ${\cal C}$ is a  $\gamma$-constacyclic code over $\Gamma$
of length $N$ if and only if ${\cal C}$ is an ideal of
the residue class ring $\Gamma[x]/\langle x^N-\gamma\rangle$, and the dual code ${\cal C}^{\bot}$ of ${\cal C}$ is a $\gamma^{-1}$-constacyclic code of length $N$ over
$\Gamma$, i.e. ${\cal C}^{\bot}$ is an ideal of the ring $\Gamma[x]/\langle
x^N-\gamma^{-1}\rangle$ (cf. [12] Propositions 2.4 and 2.5). The ring $\Gamma[x]/\langle x^N-\gamma\rangle$
is usually called the \textit{ambient ring} of $\gamma$-constacyclic codes over $\Gamma$
with length $N$. In addition,
$\mathcal{C}$ is called a \textit{simple-root constacyclic code} if
${\rm gcd}(q,N)=1$, and called a \textit{repeated-root constacyclic code} otherwise.

\par
  Let $\mathbb{F}_{q}$ be a finite field of cardinality $q$, where
$q$ is power of a prime, and denote $R=\mathbb{F}_{q}[u]/\langle u^e\rangle
=\mathbb{F}_{q}+u\mathbb{F}_{q}+\ldots+u^{e-1}\mathbb{F}_{q}$ ($u^e=0$) where $e\geq 2$. Then
$R$ is a finite chain ring. As in Dinh et al [12], if
$$\gamma=\alpha_0+\alpha_ku^k+\ldots+\alpha_{e-1}u^{e-1}$$
where $\alpha_0,\alpha_k,\ldots,\alpha_{e-1}\in \mathbb{F}_{q}$ satisfying $\alpha_0\alpha_k\neq 0$, then $\gamma$
is called a unit in $R$ to be of \textit{Type $k$}. Especially, $\gamma$
is called a unit in $R$ to be of \textit{Type $0$} if $\gamma=\alpha_0\in \mathbb{F}_{q}$. When $\gamma$ is a unit in $R$ of Type $k$, a $\gamma$-constacyclic code $\mathcal{C}$
of length $N$ over $R$ is said to be of \textit{Type $k$}. Especially,
cyclic codes and negacyclic codes are both constacyclic codes over $R$ of Type $0$.

\par
  For examples, let $e\geq 3$ and $\delta,\alpha \in \mathbb{F}_q^\times$. Then
$\delta+\alpha u^2$ is a a unit in $R$ of Type 2. Hence
$(\delta+\alpha u^2)$-constacyclic codes
form a typical subclass of the class of all Type 2 constacyclic codes over the finite chain ring $R$.

\par
  $\diamondsuit$ When $e=2$, there were a lot of literatures on linear codes,
Type $1$ constacyclic codes and some special class of Type $0$ constacyclic codes of length $N$ over rings $\mathbb{F}_{p^m}[u]/\langle u^2\rangle
=\mathbb{F}_{p^m}+u\mathbb{F}_{p^m}$ for various prime $p$ and positive integers $m$ and $N$.
See [1], [2], [13--19], [21], [25] and [27], for examples. In particular, we [3] gave an explicit representation
and a complete classification for all Type $0$ repeated-root constacyclic codes over $\mathbb{F}_{p^m}+u\mathbb{F}_{p^m}$ and their dual codes
for any prime number $p$ and positive integer $m$.

\par
  $\diamondsuit$ When $e\geq 3$, the structures for repeated-root constacyclic codes of Type 1 over $R$ had been studied by many literatures. For examples,
Kai et al. [22] investigated $(1+\lambda u)$-constacyclic codes of arbitrary length over $\mathbb{F}_p[u]/\langle u^k\rangle$, where $\lambda\in \mathbb{F}_p^\times$. Cao [4] generalized
these results to $(1+w\gamma)$-constacyclic codes of arbitrary length over an arbitrary finite
chain ring $\Gamma$, where $w$ is a unit of $\Gamma$ and $\gamma$ generates the unique maximal ideal of $\Gamma$
with nilpotency index $e\geq 2$. Hence \textsf{every Type 1 constacyclic code over any finite chain ring is
a one-generator ideal of the ambient ring}.

\par
  $\diamondsuit$ When $e\geq 3$, we known the following literatures for the structures of $(\delta+\alpha u^2)$-constacyclic codes over $R$, where $\delta,\alpha\in \mathbb{F}_{p^m}^{\times}$:
\begin{description}
\item{}
  Let $e=3$.

$\triangleright$ Sobhani [26] determined the structure of $(\delta+\alpha u^2)$-constacyclic codes
of length $p^k$ over $\mathbb{F}_{p^m}[u]/\langle u^3\rangle$.

\par
 $\triangleright$  On the basis of [7], using methods different from [26] we gave a complete description for
$(\delta+\alpha u^2)$-constacyclic codes over the ring $\mathbb{F}_{2^m}[u]/\langle u^3\rangle$ of
length $2n$ and determine explicitly the self-dual codes among them for any odd positive integer $n$ [10].

\item{}
 Let $e=4$.

 $\triangleright$ When ${\rm gcd}(q,n)=1$, in [5] for any $\delta,\alpha\in \mathbb{F}_{q}^{\times}$,
an explicit representation for all distinct $(\delta+\alpha u^2)$-constacyclic codes over the ring $\mathbb{F}_{q}[u]/\langle u^4\rangle$ of
length $n$ is given, and the dual code for each of these codes is determined. For the case of $q=2^m$ and $\delta=1$, all self-dual $(1+\alpha u^2)$-constacyclic codes over $R$ of
length $n$ are provided.

\par
 $\triangleright$ When $p=3$, in [6] an explicit representation for all distinct $(\delta+\alpha u^2)$-constacyclic codes over $\mathbb{F}_{3^m}[u]/\langle u^4\rangle$ of length
$3n$ was given, where ${\rm gcd}(3,n)=1$. Formulas for the number of all such codes and the number of codewords in
each code are provided respectively, and the dual code for each of these codes
was determined explicitly.

\par
$\triangleright$ When $p=2$, in [7] a representation and
enumeration formulas for all distinct $(\delta+\alpha u^2)$-constacyclic codes  over $\mathbb{F}_{2^m}[u]/\langle u^4\rangle$ of length $2n$ were presented explicitly, where $n$ is odd.

\par
  $\triangleright$ Mahmoodi and Sobhani [23] gave a complete classification for $(1+\alpha u^2)$-constacyclic codes of length $p^k$
over $R=\mathbb{F}_{p^m}[u]/\langle u^{4}\rangle$, where $\alpha\in \mathbb{F}_{p^m}^\times$. They determined self-dual such codes and enumerate them for the case $p=2$. Moreover, the authors discussed on Gray-maps on $R$ which preserve self-duality,
and also discuss on the images of self-dual constacyclic codes under these Gray maps.

\item{}
  Let $e=2\lambda$, where $\lambda$ is an arbitrary integer such that $\lambda\geq 2$.

$\triangleright$ Based on the results of [9], we gave a complete
description for repeated-root $(\delta+\alpha u^2)$-constacyclic codes over $\mathbb{F}_{p^m}[u]/\langle u^{2\lambda}\rangle$
for any odd prime $p$ [11].
The expressions and their derivation process for the main results in [11]
are heavily depend on that $p$ is an odd prime.

 Many methods and techniques used in [11] and [9]
can not be directly applied to the case $p=2$.
\end{description}

\par
  Motivated by those, we follow the main idea in [11], promote and develop the methods used in [7]
to determine all $(\delta+\alpha u^2)$-constacyclic codes over $R=\mathbb{F}_{2^m}[u]/\langle u^{2\lambda}\rangle$
of arbitrary even length. The
ideas and methods used in this paper are different to that used in
[23] and [26]. Therefore, we can come to clearer and more precise conclusions:
\begin{description}
\item{$\bullet$}
  Provide an explicit representation and enumeration for all distinct
$(\delta+\alpha u^2)$-constacyclic codes over $R$ of length $2^kn$ through only one theorem, for
any integer $k\geq 2$ and odd positive integer $n$. \\
Although the proof of this theorem is somewhat complicated,
the results expressed by the theorem are very clear and direct.

\item{$\bullet$}
 Obtain a clear and exact formula to count the number of all $(\delta+\alpha u^2)$-constacyclic codes over $R$ of length $2^kn$,
 and give a clear formula to count the number of codewords in each code from its generators directly.
\end{description}

\par
    The present paper is organized as follows. In Section 2,
we provide the notations
and review preparation results necessary. Then we give the main result (Theorem 2.5)
for representation and enumeration for all distinct
$(\delta+\alpha u^2)$-constacyclic code over $R$ of length $2^kn$.
In Section 3, we give an explicit representation
for a special subclass of $(\delta+\alpha u^2)$-constacyclic code over $R$ of length $2^kn$
including the particular situation of $n=1$.
Based on this, we correct a mistake for $(1+u^2)$-constacyclic codes of length $4$
over $\mathbb{F}_2[u]/\langle u^4\rangle$ listed by an example of [23]. Moreover,
we list precisely all distinct self-dual
$(1+\alpha u^2)$-constacyclic codes  of length $4$ over $\mathbb{F}_{2^m}[u]/\langle u^4\rangle$
for any $\alpha\in \mathbb{F}_{2^m}^\times$.
In Section 4, we give a proof for the main result in Section 2.
Section 5 concludes the paper.



\section{Main results}
\noindent
  In this section, we introduce the necessary notations
and review preparation results first. Then we provide the main
result on representation and enumeration for $(\delta+\alpha u^2)$-constacyclic codes over the ring $\mathbb{F}_{2^m}[u]/\langle u^{2\lambda}\rangle$ of length $2^kn$ where $k\geq 2$.

\par
   In this paper, we always assume that $m, n, \lambda, k$ are positive integers such that
$n$ is odd and $\lambda, k\geq 2$. As $\delta\in \mathbb{F}_{2^m}^\times$ and $|\mathbb{F}_{2^m}^\times|=2^m-1$,
there exists a unique element $\delta_0\in \mathbb{F}_{2^m}^\times$ such that $\delta=\delta_0^{2^k}$. This implies
$x^{2^kn}-\delta=(x^n+\delta_0)^{2^k}$ in $\mathbb{F}_{2^m}[x]$.
In this paper, we adopt the following notation.
\begin{itemize}
\item
   $R=\frac{\mathbb{F}_{2^m}[u]}{\langle u^{2\lambda}\rangle}=\mathbb{F}_{2^m}
+u\mathbb{F}_{2^m}+u^2\mathbb{F}_{2^m}+\ldots+u^{2\lambda-1}\mathbb{F}_{2^m}$ ($u^{2\lambda}=0$). Then $R$ is a finite chain ring
with the unique maximal ideal $uR$, and $2\lambda$ is the nilpotent index of $u$.

\vskip 2mm
  \item $\mathcal{A}=\frac{\mathbb{F}_{2^m}[x]}{\langle(x^{n}+\delta_0)^{2^k\lambda}\rangle}
      =\{\sum_{i=0}^{2^k\lambda n-1}a_ix^i\mid a_i\in \mathbb{F}_{2^m}, \ i=0,1,\ldots,2^k\lambda n-1\}$ in which
  the arithmetic is done modulo $(x^{n}+\delta_0)^{2^k\lambda}$. Then $\mathcal{A}$ is a finite principal ideal ring and $|\mathcal{A}|=2^{2^k\lambda mn}$.

\vskip 2mm
  \item $\mathcal{A}+u\mathcal{A}=\frac{\mathcal{A}[u]}{\langle u^2-\alpha^{-1}(x^{n}+\delta)^{2^k}\rangle}=\{\xi_0+u\xi_1\mid \xi_0,\xi_1\in \mathcal{A}\}$
 in which the operations are defined by
\begin{itemize}
\item $(\xi_0+u\xi_1)+(\eta_0+u\eta_1)=(\xi_0+\eta_0)+u(\xi_1+\eta_1)$,

\item
  $(\xi_0+u\xi_1)(\eta_0+u\eta_1)=\left(\xi_0\eta_0+\alpha^{-1}(x^{n}+\delta_0)^{2^k}\xi_1\eta_1\right)+u(\xi_0\eta_1+\xi_1\eta_0)$,
\end{itemize}
\vskip 2mm \noindent
  for all $\xi_0,\xi_1,\eta_0,\eta_1\in \mathcal{A}$. Then $\mathcal{A}$ is a subring of $\mathcal{A}+u\mathcal{A}$.
\end{itemize}

\par
 It is clear that both $\mathcal{A}+u\mathcal{A}$ and $\frac{R[x]}{\langle x^{2^kn}-(\delta+\alpha u^2)\rangle}$ are $\mathbb{F}_{2^m}$-spaces of dimension $2^{k+1}\lambda n$. Precisely, we have that
\begin{description}
\item{$\diamond$}
 $\{1,x,\ldots,x^{2^k\lambda n-1},u,ux,\ldots,ux^{2^k\lambda n-1}\}$ is an $\mathbb{F}_{2^m}$-basis of $\mathcal{A}+u\mathcal{A}$;

\item{$\diamond$}
 $\bigcup_{i=0}^{2\lambda-1}\{u^i,u^ix,\ldots,u^ix^{2^kn-1}\}$
  is an $\mathbb{F}_{2^m}$-basis of
$\frac{R[x]}{\langle x^{2^kn}-(\delta+\alpha u^2)\rangle}$.
\end{description}

\noindent
Hence there is a unique isomorphism $\Psi$ of $\mathbb{F}_{2^m}$-spaces
from $\mathcal{A}+u\mathcal{A}$ onto $\frac{R[x]}{\langle x^{2^kn}-(\delta+\alpha u^2)\rangle}$
such that
\begin{equation}
\Psi\left(x^i(x^{2^kn})^l\right)=x^i(\delta+\alpha u^2)^l
\ {\rm and} \ \Psi\left(ux^i(x^{2^kn})^l\right)=ux^i(\delta+\alpha u^2)^l,
\end{equation}
for all integers $i$ and $l$ satisfying $0\leq i\leq 2^kn-1$ and $0\leq l\leq \lambda-1$ respectively.
Moreover, by $\delta_0^{2^k}=\delta$ it follows that
\begin{description}
\item{$\diamond$}
 In the ring $\mathcal{A}+u\mathcal{A}$, we have $u^2=\alpha^{-1}(x^n+\delta_0)^{2^k}$ and $(x^n+\delta_0)^{2^k\lambda}=0$.
 These imply $x^{2^kn}=\delta+\alpha u^2$ and $u^{2\lambda}=(\alpha^{-1}(x^n+\delta_0)^{2^k})^\lambda=0$, respectively.

\item{$\diamond$}
 In the ring $\frac{R[x]}{\langle x^{2^kn}-(\delta+\alpha u^2)\rangle}$, we have
 $x^{2^kn}-(\delta+\alpha u^2)=0$ and $u^{2\lambda}=0$. These imply $(x^n+\delta_0)^{2^k\lambda}=(x^{2^kn}-\delta)^\lambda
 =(\alpha u^2)^\lambda=0$.
\end{description}

\noindent
  From these and by an argument similar to the proof of Theorem 2.1 in [11], one can
easily verify the following conclusion.

\vskip 3mm \noindent
   {\bf Lemma 2.1} \textit{Using the notations above, $\Psi$ is a ring isomorphism from
$\mathcal{A}+u\mathcal{A}$ onto $R[x]/\langle x^{2^kn}-(\delta+\alpha u^2)\rangle$}.

\vskip 3mm \par
  From this lemma and by Equation (1), we deduce the following:
\begin{equation}
\Psi\left((x^n+\delta_0)^{i+l\cdot 2^kn}\right)=\alpha^lu^{2l}(x^n+\delta_0)^i,
\ \forall 0\leq i\leq 2^kn-1, \ \forall 0\leq l\leq \lambda-1.
\end{equation}

\noindent
In the rest of this paper, we usually identify
$R[x]/\langle x^{2^kn}-(\delta+\alpha u^2)\rangle$ with
$\mathcal{A}+u\mathcal{A}$ under the ring isomorphism $\Psi$ determined
by Equations (1) and (2), unless otherwise stated.
Hence
in order to determine all $(\delta+\alpha u^2)$-constacyclic codes over $R$ of length $2^kn$,
it is sufficient to give all
ideals of the ring $\mathcal{A}+u\mathcal{A}$.

\par
  To determine all
ideals of the ring $\mathcal{A}+u\mathcal{A}$, we need to study
the structure of  the ring $\mathcal{A}$ first.
As $n$ is odd, there are pairwise coprime monic
irreducible polynomials $f_1(x),\ldots, f_r(x)$ in $\mathbb{F}_{2^m}[x]$ such that $x^{n}-\delta_0=f_1(x)\ldots f_r(x)$. This
implies
$$(x^{2^kn}-\delta)^\lambda=(x^n+\delta_0)^{2^k\lambda}=f_1(x)^{2^k\lambda}\ldots f_r(x)^{2^k\lambda}.$$
For any integer $j$, $1\leq j\leq r$, we assume ${\rm deg}(f_j(x))=d_j$ and denote
$F_j(x)=\frac{x^{n}-\delta_0}{f_j(x)}$.
Then $F_j(x)^{2^k\lambda}=\frac{(x^{2^kn}-\delta)^\lambda}{f_j(x)^{2^k\lambda}}$ and ${\rm gcd}(F_j(x),f_j(x))=1$. These imply
$$
 (x^{2^kn}-\delta)^\lambda=F_j(x)^{2^k\lambda}f_j(x)^{2^k\lambda} \ {\rm and} \ {\rm gcd}(F_j(x)^{2^k\lambda},f_j(x)^{2^k\lambda})=1.
$$
Hence there exist $g_j(x),h_j(x)\in \mathbb{F}_{2^m}[x]$ such that
$$
g_j(x)F_j(x)^{\lambda p^k}+h_j(x)f_j(x)^{\lambda p^k}=1.
$$
\noindent
As in [11] for odd prime $p$, we adopt the following notation where $j$ be an integer satisfying $1\leq j\leq r$.
\begin{itemize}
 \item
 Let $\varepsilon_j(x)\in \mathcal{A}$ be defined by
$$\varepsilon_j(x)\equiv g_j(x)F_j(x)^{2^k\lambda}=1-h_j(x)f_j(x)^{2^k\lambda} \ ({\rm mod} \ (x^{n}+\delta_0)^{2^k\lambda}).$$

 \item
 $\mathcal{K}_j=\frac{\mathbb{F}_{2^m}[x]}{\langle f_j(x)^{2^k\lambda}\rangle}=\{\sum_{i=0}^{2^k\lambda d_j-1}a_ix^i\mid
a_i\in \mathbb{F}_{2^m}, 0\leq i<2^k\lambda d_j\}$ in which the arithmetics are done modulo $f_j(x)^{2^k\lambda}$.

\vskip 2mm
 \item
 $\mathcal{F}_j=\frac{\mathbb{F}_{2^m}[x]}{\langle f_j(x)\rangle}=\{\sum_{i=0}^{d_j-1}a_ix^i\mid
a_i\in \mathbb{F}_{2^m}, 0\leq i<d_j\}$ in which the arithmetics are done modulo $f_j(x)$.
Then $\mathcal{F}_j$ is an extension field of $\mathbb{F}_{2^m}$ with $2^{md_j}$ elements.
\end{itemize}

\noindent
  {\bf Remark} $\mathcal{F}_j$ is a finite field in which the arithmetic is done
modulo $f_j(x)$, $\mathcal{K}_j$ is a finite ring in which the arithmetic is done
modulo $f_j(x)^{2^k\lambda}$ and $\mathcal{A}$ is a principal ideal ring in which the arithmetic is done
modulo $(x^{n}+\delta_0)^{2^k\lambda}$. In this paper, we adopt the following points of view:
$$\mathcal{F}_j\subseteq \mathcal{K}_j\subseteq \mathcal{A} \ {\rm as} \ {\rm sets}.$$
It is worth noting that $\mathcal{F}_j$ is not a subfield of $\mathcal{K}_j$ and $\mathcal{K}_j$ is not a subring of $\mathcal{A}$
when $n\geq 2$.

\vskip 3mm \par
  Then from Chinese remainder theorem for commutative rings, we deduce the following lemma about the structure and
properties of the ring $\mathcal{A}$.

\vskip 3mm
\noindent
  {\bf Lemma 2.2} \textit{Using the notations above, we have the following  decomposition
via idempotents}:

\vskip 2mm \par
  (i) \textit{$\varepsilon_1(x)+\ldots+\varepsilon_r(x)=1$, $\varepsilon_j(x)^2=\varepsilon_j(x)$
and $\varepsilon_j(x)\varepsilon_l(x)=0$  in the ring $\mathcal{A}$ for all $1\leq j\neq l\leq r$}.

\vskip 2mm \par
  (ii) \textit{We regard $\mathcal{K}_j$ as a subset of $\mathcal{A}$ for all $j$. Then}
$$\mathcal{A}=\varepsilon_1(x)\mathcal{K}_1\oplus\varepsilon_2(x)\mathcal{K}_2\oplus
\ldots\oplus\varepsilon_r(x)\mathcal{K}_r \ ({\rm mod} \ (x^n+\delta_0)^{2^k\lambda}),$$
\textit{where $\varepsilon_j(x)\mathcal{K}_j=\{\varepsilon_j(x)a_j(x)\mid a_j(x)\in \mathcal{K}_j\}$ for all $j=1,\ldots, r$}.

\vskip 3mm\par
  For the ring $\mathcal{K}_j$, where $1\leq j\leq r$, we know the following conclusion.

\vskip 3mm\noindent
  {\bf Lemma 2.3}
  (cf. [8] Example 2.1) \textit{The ring $\mathcal{K}_j$ have the following properties}:

\begin{description}
\item{(i)}
  \textit{$\mathcal{K}_j$ is a finite chain ring, $f_j(x)$ generates the unique
maximal ideal $\langle f_j(x)\rangle$ of $\mathcal{K}_j$, $2^k\lambda$ is the nilpotency index of $f_j(x)$ and the residue class field of $\mathcal{K}_j$ modulo $\langle f_j(x)\rangle$ is $\mathcal{K}_j/\langle f_j(x)\rangle\cong \mathcal{F}_{j}$}.

\item{(ii)}
 \textit{Every element $\xi$ of $\mathcal{K}_j$ has a unique $f_j(x)$-adic expansion}:
$$\xi=b_0(x)+b_1(x)f_j(x)+\ldots+ b_{2^k\lambda-1}f_j(x)^{2^k\lambda-1},$$
\textit{where $\ b_0(x), b_1(x),\ldots, b_{2^k\lambda-1}\in \mathcal{F}_j$. Moreover,
$\xi$ is invertible in $\mathcal{K}_j$ if and only if $b_0(x)\neq 0$. Here regard $\mathcal{F}_j$ as a subset of $\mathcal{K}_j$}.

\item{(iii)}
   \textit{All distinct $2^k\lambda+1$ ideals of $\mathcal{K}_j$ are given by:
   $$\langle f_j(x)^l\rangle=f_j(x)^l\mathcal{K}_j, \ 0\leq l\leq 2^k\lambda.$$
   Moreover, $|\langle f_j(x)^l\rangle|=2^{md_j(2^k\lambda-l)}$ for $l=0,1,\ldots,2^k\lambda$}.

\item{(iv)}
  \textit{Let $1\leq l\leq 2^k\lambda$. Then $|\mathcal{K}_j/\langle f_j(x)^l\rangle|=2^{md_jl}$. Precisely, we have}
$$\mathcal{K}_j/\langle f_j(x)^l\rangle=\{\sum_{i=0}^{l-1}b_i(x)f(x)^i\mid b_i(x)\in \mathcal{F}_j, \ i=0,1,\ldots,l-1\}.$$
\end{description}

\noindent
  {\bf Remark} For any integer $l$, $1\leq l\leq 2^k\lambda-1$, by Lemma 2.3(iv)
we can identify $\mathcal{K}_j/\langle f_j(x)^l\rangle$ with $\frac{\mathbb{F}_{2^m}[x]}{\langle f_j(x)^l\rangle}$ up to a natural
ring isomorphism. We will take this view in the rest of this paper.
  Then for any
$0\leq l\leq t\leq 2^k\lambda-1$, we stipulate
\begin{eqnarray*}
f_j(x)^l\cdot \frac{\mathbb{F}_{2^m}[x]}{\langle f_j(x)^t\rangle}
  &=&f_j(x)^l(\mathcal{K}_j/\langle f_j(x)^t\rangle)\\
  &=&\{\sum_{i=l}^{t-1}b_i(x)f(x)^i\mid b_i(x)\in \mathcal{F}_j, \ i=l,\ldots,t-1\}.
\end{eqnarray*}
Hence $|f_j(x)^l(\mathcal{K}_j/\langle f_j(x)^t\rangle)|=2^{md_j(t-l)}$, where we set
$f_j(x)^l(\mathcal{K}_j/\langle f_j(x)^l\rangle)$ $=\{0\}$ for convenience.

\vskip 3mm\par
  Then we study the structure of  the ring $\mathcal{A}+u\mathcal{A}$. To do this,
we introduce the following notation.
\begin{itemize}
 \item
 Let $\alpha_0\in \mathbb{F}_{2^m}^\times$ satisfying $\alpha_0^2=\alpha^{-1}$.

\item
 Let $\omega_j\in \mathcal{K}_j$ be defined by $\omega_j=\alpha_0F_j(x)^{2^{k-1}} \ ({\rm mod} \ f_j(x)^{2^k\lambda})$.

\vskip 2mm
  \item $\mathcal{K}_j+u\mathcal{K}_j=\frac{\mathcal{K}_j[u]}{\langle u^2-\omega_j^2f_j(x)^{2^k}\rangle}=\{\xi_0+u\xi_1\mid \xi_0,\xi_1\in \mathcal{K}_j\}$
 in which the operations are defined by
\begin{itemize}
\item $(\xi_0+u\xi_1)+(\eta_0+u\eta_1)=(\xi_0+\eta_0)+u(\xi_1+\eta_1)$,

\item
  $(\xi_0+u\xi_1)(\eta_0+u\eta_1)=\left(\xi_0\eta_0+\omega_j^2f_j(x)^{2^k}\xi_1\eta_1\right)+u(\xi_0\eta_1+\xi_1\eta_0)$,
\end{itemize}
\vskip 2mm \noindent
  for all $\xi_0,\xi_1,\eta_0,\eta_1\in \mathcal{K}_j$. Then $\mathcal{K}_j$ is a subring of $\mathcal{K}_j+u\mathcal{K}_j$.
\end{itemize}

\noindent
  {\bf Lemma 2.4} \textit{Let $1\leq j\leq r$. Then we have the following conclusions}.
\begin{description}
\item{(i)}
  \textit{The element $\omega_j$ is invertible in the ring $\mathcal{K}_j$ and satisfies
$$\omega_j^2f_j(x)^{2^k}\equiv \alpha^{-1}(x^{n}+\delta_0)^{2^k} \ ({\rm mod} \ \ f_j(x)^{2^k\lambda}).$$
Hence $\alpha^{-1}(x^{n}+\delta_0)^{2^k}=\omega_j^2f_j(x)^{2^k}$ in the ring
$\mathcal{K}_j$}.

\item{(ii)}
\textit{We regard $\mathcal{K}_j+u\mathcal{K}_j$ as a subset of $\mathcal{A}+u\mathcal{A}$ for all $j$. Then}
$$\mathcal{A}+u\mathcal{A}=\bigoplus_{j=1}^r\varepsilon_j(x)(\mathcal{K}_j+u\mathcal{K}_j)
 =\sum_{j=1}^r\varepsilon_j(x)(\mathcal{K}_j+u\mathcal{K}_j) \ ({\rm mod} \ (x^n+\delta_0)^{2^k\lambda}),$$
\textit{where $\varepsilon_j(x)(\mathcal{K}_j+u\mathcal{K}_j)=\{\varepsilon_j(x)a_j(x)+u\varepsilon_j(x)b_j(x)\mid a_j(x),b_j(x)\in \mathcal{K}_j\}$}.
\end{description}

\noindent
 {\bf Proof}. (i) Since $\omega_j\in \mathcal{K}_j$ satisfying $\omega_j\equiv\alpha_0F_j(x)^{2^{k-1}}$ (mod $f_j(x)^{2^k\lambda}$)
and ${\rm gcd}(F_j(x),f_j(x))=1$, we conclude that ${\rm gcd}(\omega_j,f_j(x)^{2^k\lambda})=1$
as polynomials in $\mathbb{F}_{2^m}[x]$. This implies that $\omega_j$ is an invertible element of the ring $\mathcal{K}_j$.
Then from $(x^{n}+\delta_0)^{2^k}=f_1(x)^{2^k}\ldots f_r(x)^{2^k}$ and $F_j(x)^{2^k}=\frac{(x^{n}+\delta_0)^{2^k}}{f_j(x)^{2^k}}$, we deduce that
\begin{eqnarray*}
\omega_j^2f_j(x)^{2^k} &\equiv& (\alpha_0F_j(x)^{2^{k-1}})^2f_j(x)^{2^k}=\alpha^{-1}F_j(x)^{2^{k}}f_j(x)^{2^k} \\
&\equiv & \alpha^{-1}(x^{n}+\delta_0)^{2^k} \ ({\rm mod} \ \ f_j(x)^{2^k\lambda}).
\end{eqnarray*}
This implies $\alpha^{-1}(x^{n}+\delta_0)^{2^k}=\omega_j^2f_j(x)^{2^k}$ in $\mathcal{K}_j$.

\par
  (ii) By Lemma 2.2 and the conclusion of (i), it follows that
\begin{eqnarray*}
\mathcal{A}+u\mathcal{A}
  &=&\frac{\mathcal{A}[u]}{\langle u^2-\alpha^{-1}(x^{n}+\delta_0)^{2^k}\rangle}
 =\frac{\left(\bigoplus_{j=1}^r\varepsilon_j(x)\mathcal{K}_j\right)[u]}{\langle u^2-\alpha^{-1}(x^{n}+\delta_0)^{2^k}\rangle}\\
 &=& \bigoplus_{j=1}^r\varepsilon_j(x)\cdot \frac{\mathcal{K}_j[u]}{\langle u^2-\alpha^{-1}(x^{n}+\delta_0)^{2^k}\rangle}\\
 &=& \bigoplus_{j=1}^r\varepsilon_j(x)\cdot \frac{\mathcal{K}_j[u]}{\langle u^2-\omega_j^2f_j(x)^{2^k}\rangle}.
  \ ({\rm mod} \ (x^n+\delta_0)^{2^k\lambda}).
\end{eqnarray*}
This implies $\mathcal{A}+u\mathcal{A}=\bigoplus_{j=1}^r\varepsilon_j(x)(\mathcal{K}_j+u\mathcal{K}_j)
 =\sum_{j=1}^r\varepsilon_j(x)(\mathcal{K}_j+u\mathcal{K}_j)$.
\hfill $\Box$

\vskip 3mm
\par
  Finally, we list all $(\delta+\alpha u^2)$-constacyclic
codes over $R$ of length $2^kn$ by the following theorem. Its proof
will be given in Section 4.

\vskip 3mm
\noindent
  {\bf Theorem 2.5} \textit{For any integers $j,l$: $1\leq j\leq r$ and $0\leq l\leq 2^{k-1}\lambda$, let
$$\frac{\mathbb{F}_{2^m}[x]}{\langle f_j(x)^l\rangle}
=\left\{\sum_{i=0}^{l-1}a_i(x)f_j(x)^i\mid a_0(x),a_1(x),\ldots,a_{l-1}(x)\in \mathcal{F}_{j}\right\},$$
 Especially, we have $\frac{\mathbb{F}_{2^m}[x]}{\langle f_j(x)^0\rangle}=\{0\}$
and $\frac{\mathbb{F}_{2^m}[x]}{\langle f_j(x)\rangle}=\mathcal{F}_{j}$. Then all distinct
$(\delta+\alpha u^2)$-constacyclic
codes over $R$ of length $2^kn$, as ideals of the ring $\mathcal{A}+u\mathcal{A}$, are given by
$$\mathcal{C}=\bigoplus_{j=1}^r\varepsilon_j(x)C_j,$$
where $C_j$ is an ideal
of the ring $\mathcal{K}_j+u\mathcal{K}_j$ listed by the following three cases}.
\begin{description}
   \item{(I)}
   $\sum_{s=0}^{2^{k}\lambda-1}2^{md_j(2^{k-1}\lambda-\lceil\frac{s}{2}\rceil)}$ \textit{ideals given by the following two subcases}:

\begin{description}
   \item{{\bf 1.}}
   $C_j=\left\langle \omega_jf_j(x)^{2^{k-1}+s}
+f_j(x)^{2^{k-1}\lambda+\lceil \frac{s}{2}\rceil}h(x)+u f_j(x)^s\right\rangle$ \textit{with} $|C|=2^{md_j(2^{k}\lambda-s)}$,
 \textit{where $h(x)\in  \frac{\mathbb{F}_{2^m}[x]}{\langle f_j(x)^{2^{k-1}\lambda-\lceil \frac{s}{2}\rceil}\rangle}$ and $0\leq s\leq 2^{k-1}\lambda-1$}.

  \item{{\bf 2.}}
   $C_j=\left\langle f_j(x)^{2^{k-1}\lambda+\lceil \frac{s}{2}\rceil}h(x)+u f_j(x)^s\right\rangle$ \textit{with} $|C|=2^{md_j(2^{k}\lambda-s)}$, \\ \textit{where $h(x)\in  \frac{\mathbb{F}_{2^m}[x]}{\langle f_j(x)^{2^{k-1}\lambda-\lceil \frac{s}{2}\rceil}\rangle}$ and $2^{k-1}\lambda\leq s\leq 2^{k}\lambda-1$}.

\end{description}

\item{(II)}
 $2^{k}\lambda+1$ \textit{ideals}:
 \begin{description}
   \item{{\bf 3.}}
   \textit{$C_j=\left\langle f_j(x)^s\right\rangle$ with $|C|=2^{md_j(2^{k+1}\lambda-2s)}$, where $0\leq s\leq 2^{k}\lambda$}.
 \end{description}

\item{(III)}
 $\sum_{t=1}^{2^{k}\lambda-1}(2^{k}\lambda-t)2^{md_j\lfloor\frac{t}{2}\rfloor}$
 \textit{ideals given by the following three subcases}:

\begin{description}
   \item{{\bf 4.}}
   \textit{$C_j=\left\langle u f_j(x)^s, f_j(x)^{s+1}\right\rangle$ with $|C|=2^{md_j(2^{k+1}\lambda-2s-1)}$, \\
   where $0\leq s\leq 2^{k}\lambda-2$}.

\item{{\bf 5.}}
  \textit{$C_j=\left\langle f_j(x)^{s+\lceil \frac{t}{2}\rceil}h(x)+uf_j(x)^s, f_j(x)^{s+t}\right\rangle$ \\
  with $|C|=2^{md_j(2^{k}\lambda-2s-t)}$,
  where $h(x)\in\frac{\mathbb{F}_{2^m}[x]}{\langle f_j(x)^{\lfloor \frac{t}{2}\rfloor}\rangle}$, \\
   $0\leq s\leq 2^{k}\lambda-1-t$ and $2\leq t\leq 2^k$.}

\item{{\bf 6.}}
  \textit{$C_j=\left\langle \omega_jf_j(x)^{2^{k-1}+s}+f_j(x)^{s+\lceil \frac{t}{2}\rceil}h(x)+uf_j(x)^s, f_j(x)^{s+t}\right\rangle$ \\ with $|C|=2^{md_j(2^{k+1}\lambda-2s-t)}$,
  where $h(x)\in\frac{\mathbb{F}_{2^m}[x]}{\langle f_j(x)^{\lfloor \frac{t}{2}\rfloor}\rangle}$, \\
   $0\leq s\leq 2^{k}\lambda-1-t$ and $2^k+1\leq t\leq 2^{k}\lambda-1$.}
\end{description}
   \textit{In this case, the number of codewords in $\mathcal{C}$ is
   $|\mathcal{C}|=\prod_{j=1}^r|C_j|$}.
\end{description}

\par
  \textit{Moreover, the number of
all $(\delta+\alpha u^2)$-constacyclic
codes over $R$ of length $2^kn$ is equal to
$\prod_{j=1}^rN_{(2^m,d_j,2^{k}\lambda)},$
where}
\begin{eqnarray*}
N_{(2^m,d_j,2^{k}\lambda)}
 &=& \sum_{i=0}^{2^{k-1}\lambda}(1+4i)2^{md_j(2^{k-1}\lambda-i)m}\\
 &=& \frac{(2^{md_j}+3)2^{md_j(2^{k-1}\lambda+1)}-2^{md_j}(2^{k+1}\lambda+5)+2^{k+1}\lambda+1}{(2^{md_j}-1)^2}.
\end{eqnarray*}
\textit{is the number of ideals in $\mathcal{K}_j+u\mathcal{K}_j$ for all $j=1,\ldots,r$}.

\vskip 3mm\par
   Specifically, one can easily give an explicit representation for
all distinct $(\delta+\alpha u^2)$-constacyclic codes over $R$ of length $2^kn$, as
ideals of the ring $\frac{R[x]}{\langle x^{2^kn}-(\delta+\alpha u^2)\rangle}$, from
Theorem 2.5 by replacing each element in $\mathcal{C}$ with the
substitutions determined by Equations (1) and (2).

\par
  Similarly, as Corollary 3.8 of [11] one can prove the following conclusion.

\vskip 3mm
\noindent
  {\bf Corollary 2.6} \textit{Every $(\delta+\alpha u^2)$-constacyclic code over $R$ of length $2^kn$ can be
generated by at most $2$ polynomials in the ring $\frac{R[x]}{\langle x^{2^kn}-(\delta+\alpha u^2)\rangle}$}.


\section{A subclass of $(\delta+\alpha u^2)$-constacyclic codes over $R$} \label{}
\noindent
In this section, let $x^n+\delta_0$ be an irreducible polynomial in
$\mathbb{F}_{2^m}[x]$ and $\delta=\delta_0^{2^k}$. We consider $(\delta+\alpha u^2)$-constacyclic codes over $R$
of length $2^kn$.
In this case, we have $r=1$, $f_1(x)=x^n+\delta_0$, $\varepsilon_1(x)=1$, $d_1={\rm deg}(x^n-\delta_0)=n$,
$\omega_1=\alpha_0$ where $\alpha_0^2=\alpha^{-1}$, and
\begin{itemize}
 \item
  $\mathcal{K}_1=\mathcal{A}=\frac{\mathbb{F}_{2^m}[x]}{\langle (x^{n}+\delta_0)^{2^k\lambda}\rangle}$;

 \item
  $\mathcal{F}_1=\frac{\mathbb{F}_{2^m}[x]}{\langle x^{n}+\delta_0\rangle}=\{\sum_{i=0}^{n-1}a_ix^i\mid a_0,a_1,\ldots,a_{n-1}\in \mathbb{F}_{2^m}\}$ and $|\mathcal{F}_1|=2^{mn}$.
\end{itemize}

\noindent
  From these and by Theorem 2.5, we deduce the following conclusion.

\vskip 3mm \noindent
  {\bf Theorem 3.1} \textit{Let $x^n+\delta_0$ be an irreducible polynomial in
$\mathbb{F}_{2^m}[x]$ and $\delta=\delta_0^{2^k}$. Then all distinct
$(\delta+\alpha u^2)$-constacyclic codes over $R$ of length $2^kn$ are listed by the following three cases}.
\begin{description}
   \item{(I)}
   $\sum_{s=0}^{2^{k}\lambda-1}2^{mn(2^{k-1}\lambda-\lceil\frac{s}{2}\rceil)}$ \textit{codes given by the following two subcases}:

\begin{description}
   \item{{\bf 1.}}
   $\mathcal{C}=\langle \alpha_0(x^n+\delta_0)^{2^{k-1}+s}
+(x^n+\delta_0)^{2^{k-1}\lambda+\lceil \frac{s}{2}\rceil}h(x)+u (x^n+\delta_0)^s\rangle$ \\
 \textit{with} $|C|=2^{mn(2^{k}\lambda-s)}$,
 \textit{where $h(x)\in  \frac{\mathbb{F}_{2^m}[x]}{\langle (x^n+\delta_0)^{2^{k-1}\lambda-\lceil \frac{s}{2}\rceil}\rangle}$ \\
 and $0\leq s\leq 2^{k-1}\lambda-1$}.

  \item{{\bf 2.}}
   $\mathcal{C}=\langle (x^n+\delta_0)^{2^{k-1}\lambda+\lceil \frac{s}{2}\rceil}h(x)+u (x^n+\delta_0)^s\rangle$ \textit{with} $|C|=2^{mn(2^{k}\lambda-s)}$, \\ \textit{where $h(x)\in  \frac{\mathbb{F}_{2^m}[x]}{\langle (x^n+\delta_0)^{2^{k-1}\lambda-\lceil \frac{s}{2}\rceil}\rangle}$ and $2^{k-1}\lambda\leq s\leq 2^{k}\lambda-1$}.
\end{description}

\item{(II)}
 $2^{k}\lambda+1$ \textit{codes}:
 \begin{description}
   \item{{\bf 3.}}
   \textit{$\mathcal{C}=\langle (x^n+\delta_0)^s\rangle$ with $|C|=2^{mn(2^{k+1}\lambda-2s)}$, where $0\leq s\leq 2^{k}\lambda$}.
 \end{description}

\item{(III)}
 $\sum_{t=1}^{2^{k}\lambda-1}(2^{k}\lambda-t)2^{mn\lfloor\frac{t}{2}\rfloor}$
 \textit{codes given by the following three subcases}:

\begin{description}
   \item{{\bf 4.}}
   \textit{$\mathcal{C}=\langle u (x^n+\delta_0)^s, (x^n+\delta_0)^{s+1}\rangle$ \\
   with $|C|=2^{mn(2^{k+1}\lambda-2s-1)}$,
   where $0\leq s\leq 2^{k}\lambda-2$}.

\item{{\bf 5.}}
  \textit{$\mathcal{C}=\langle (x^n+\delta_0)^{s+\lceil \frac{t}{2}\rceil}h(x)+u(x^n+\delta_0)^s, (x^n+\delta_0)^{s+t}\rangle$ \\
  with $|C|=2^{mn(2^{k}\lambda-2s-t)}$,
  where $h(x)\in\frac{\mathbb{F}_{2^m}[x]}{\langle (x^n+\delta_0)^{\lfloor \frac{t}{2}\rfloor}\rangle}$, \\
   $0\leq s\leq 2^{k}\lambda-1-t$ and $2\leq t\leq 2^k$.}

\item{{\bf 6.}}
  \textit{$\mathcal{C}=\langle \alpha_0(x^n+\delta_0)^{2^{k-1}+s}+(x^n+\delta_0)^{s+\lceil \frac{t}{2}\rceil}h(x)+u(x^n+\delta_0)^s, (x^n+\delta_0)^{s+t}\rangle$ \\
  with $|C|=2^{mn(2^{k+1}\lambda-2s-t)}$,
  where $h(x)\in\frac{\mathbb{F}_{2^m}[x]}{\langle (x^n+\delta_0)^{\lfloor \frac{t}{2}\rfloor}\rangle}$, \\
   $0\leq s\leq 2^{k}\lambda-1-t$ and $2^k+1\leq t\leq 2^{k}\lambda-1$.}
\end{description}
\end{description}

\par
  \textit{In this case, the number of codes of
all $(\delta+\alpha u^2)$-constacyclic
codes over $R$ of length $2^kn$ is equal to}
\begin{eqnarray*}
N_{(2^m,n,2^{k}\lambda)}
 &=& \sum_{i=0}^{2^{k-1}\lambda}(1+4i)2^{mn(2^{k-1}\lambda-i)m}\\
 &=& \frac{(2^{mn}+3)2^{mn(2^{k-1}\lambda+1)}-2^{mn}(2^{k+1}\lambda+5)+2^{k+1}\lambda+1}{(2^{mn}-1)^2}.
\end{eqnarray*}

\vskip 3mm\par
  Especially, set $n=1$ in Theorem 3.1. We obtain the following corollary.

\vskip 3mm
\noindent
  {\bf Corollary 3.2} \textit{All distinct $(\delta+\alpha u^2)$-constacyclic codes over $R$ of length $2^k$ are listed by
the following three cases}.
\begin{description}
   \item{(I)}
   $\sum_{s=0}^{2^{k}\lambda-1}2^{m(2^{k-1}\lambda-\lceil\frac{s}{2}\rceil)}$ \textit{codes given by the following two subcases}:

\begin{description}
   \item{{\bf 1.}}
   $C=\langle \alpha_0(x+\delta_0)^{2^{k-1}+s}
+(x+\delta_0)^{2^{k-1}\lambda+\lceil \frac{s}{2}\rceil}h(x)+u (x+\delta_0)^s\rangle$ \textit{with} $|C|=2^{m(2^{k}\lambda-s)}$,
 \textit{where $h(x)\in  \frac{\mathbb{F}_{2^m}[x]}{\langle (x+\delta_0)^{2^{k-1}\lambda-\lceil \frac{s}{2}\rceil}\rangle}$ and $0\leq s\leq 2^{k-1}\lambda-1$}.

  \item{{\bf 2.}}
   $C=\langle (x+\delta_0)^{2^{k-1}\lambda+\lceil \frac{s}{2}\rceil}h(x)+u (x+\delta_0)^s\rangle$ \textit{with} $|C|=2^{m(2^{k}\lambda-s)}$, \\ \textit{where $h(x)\in  \frac{\mathbb{F}_{2^m}[x]}{\langle (x+\delta_0)^{2^{k-1}\lambda-\lceil \frac{s}{2}\rceil}\rangle}$ and $2^{k-1}\lambda\leq s\leq 2^{k}\lambda-1$}.

\end{description}

\item{(II)}
 $2^{k}\lambda+1$ \textit{codes}:
 \begin{description}
   \item{{\bf 3.}}
   \textit{$C=\langle (x+\delta_0)^s\rangle$ with $|C|=2^{m(2^{k+1}\lambda-2s)}$, where $0\leq s\leq 2^{k}\lambda$}.
 \end{description}

\item{(III)}
 $\sum_{t=1}^{2^{k}\lambda-1}(2^{k}\lambda-t)2^{m\lfloor\frac{t}{2}\rfloor}$
 \textit{codes given by one of the following three subcases}:

\begin{description}
   \item{{\bf 4.}}
   \textit{$C=\langle u (x+\delta_0)^s, (x+\delta_0)^{s+1}\rangle$ with $|C|=2^{m(2^{k+1}\lambda-2s-1)}$, \\
   where $0\leq s\leq 2^{k}\lambda-2$}.

\item{{\bf 5.}}
  \textit{$C=\langle (x+\delta_0)^{s+\lceil \frac{t}{2}\rceil}h(x)+u(x+\delta_0)^s, (x+\delta_0)^{s+t}\rangle$ with $|C|=2^{m(2^{k+1}\lambda-2s-t)}$, \\
  where $h(x)\in\frac{\mathbb{F}_{2^m}[x]}{\langle (x+\delta_0)^{\lfloor \frac{t}{2}\rfloor}\rangle}$,
   $0\leq s\leq 2^{k}\lambda-1-t$ and $2\leq t\leq 2^k$.}

\item{{\bf 6.}}
  \textit{$C=\langle \alpha_0(x+\delta_0)^{2^{k-1}+s}+(x+\delta_0)^{s+\lceil \frac{t}{2}\rceil}h(x)+u(x+\delta_0)^s, (x+\delta_0)^{s+t}\rangle$ with $|C|=2^{m(2^{k+1}\lambda-2s-t)}$,
  where $h(x)\in\frac{\mathbb{F}_{2^m}[x]}{\langle (x+\delta_0)^{\lfloor \frac{t}{2}\rfloor}\rangle}$,
   $0\leq s\leq 2^{k}\lambda-1-t$ and $2^k+1\leq t\leq 2^{k}\lambda-1$.}
\end{description}
\end{description}

\par
  \textit{Therefore, the number of ideals of the ring $\mathcal{A}+u\mathcal{A}$
is equal to}
\begin{eqnarray*}
N_{(2^m,1,2^{k}\lambda)}
 &=& \sum_{i=0}^{2^{k-1}\lambda}(1+4i)2^{m(2^{k-1}\lambda-i)m}\\
 &=& \frac{(2^{m}+3)2^{(2^{k-1}\lambda+1)m}-2^{m}(2^{k+}\lambda+5)+2^{k+1}\lambda+1}{(2^{m}-1)^2}.
\end{eqnarray*}

\par
  \vskip 3mm\par
  As the end of this section, we list all $(\delta+\alpha u^2)$-constacyclic code over
$R=\frac{\mathbb{F}_{2^m}[u]}{\langle u^4\rangle}$ of length $4$,
where $\delta,\alpha, \delta_0,\alpha_0\in \mathbb{F}_{2^m}^\times$ satisfying
$\delta_0^4=\delta$ and $\delta_0^2=\alpha^{-1}$ respectively. In this case, we have
have the following:
\begin{description}
\item{$\checkmark$}
  $\Psi((x+\delta_0)^4)=\alpha u^2=\alpha_0^{-2} u^2$.

\item{$\checkmark$}
 $\frac{\mathbb{F}_{2^m}[x]}{\langle (x+\delta_0)^{l}\rangle}=\{\alpha h(x)\mid
h(x)\in \frac{\mathbb{F}_{2^m}[x]}{\langle (x+\delta_0)^{l}\rangle}\}$ for all $l=0,1,2,3,4$.

\item{$\checkmark$}
 $\Psi((x+\delta_0)^{3}h(x))=(x+\delta_0)^{3}h_0+u^2h_1$, where $h_1=\alpha \widetilde{h}_1$,
 for all $h(x)=h_0+\widetilde{h}_1(x+\delta_0)\in \frac{\mathbb{F}_{2^m}[x]}{\langle (x+\delta_0)^{2}\rangle}$
 with $h_0, \widetilde{h}_1\in \mathbb{F}_{2^m}$.

\item{$\checkmark$}
 $\Psi((x+\delta_0)^{3}h(x))=(x+\delta_0)^{3}h_0+u^2h_1+u^2(x+\delta_0)h_2$, where $h_1=\alpha \widetilde{h}_1$
 and $h_2=\alpha \widetilde{h}_2$,
 for all $h(x)=h_0+\widetilde{h}_1(x+\delta_0)+\widetilde{h}_2(x+\delta_0)^2\in \frac{\mathbb{F}_{2^m}[x]}{\langle (x+\delta_0)^{3}\rangle}$
 with $h_0, \widetilde{h}_1, \widetilde{h}_2\in \mathbb{F}_{2^m}$.
\end{description}

\noindent
  First, by Corollary 3.2
the number of $(\delta+\alpha u^2)$-constacyclic code over
$R=\frac{\mathbb{F}_{2^m}[u]}{\langle u^4\rangle}$ of length $4$ is equal to
$$N_{(2^m,1,2^{2}\cdot 2)}=N_{(2^m,1,2^{3})}=17+13\cdot 2^m+9\cdot (2^m)^2+5\cdot(2^m)^3+(2^m)^4.$$
Then we list all these codes by the following corollary.

\vskip 3mm
\noindent
  {\bf Corollary 3.3} \textit{Denote
$y=x+\delta_0$ and
$\mathcal{F}_l=\frac{\mathbb{F}_{2^m}[x]}{\langle (x+\delta_0)^{l}\rangle}$
for $l=2,3,4$.
Then all distinct $(\delta+\alpha u^2)$-constacyclic codes over $\frac{\mathbb{F}_{2^m}[u]}{\langle u^4\rangle}$ of length $4$ are given by the following
tables}:
{\footnotesize
\begin{center}
\begin{tabular}{l|llll}\hline
Case & $\mathcal{L}$ &  $\mathcal{C}$ & & $|\mathcal{C}|$ \\ \hline
{\bf 1.} & $(2^m)^4$ & $\langle \alpha_0y^2+u^2 \cdot h(x)+u\rangle$, & $h(x)\in \mathcal{F}_4$ & $2^{8m}$ \\
         &  $(2^m)^3$ & $\langle \alpha_0y^3+u^2\cdot yh(x)+u y\rangle$, & $h(x)\in \mathcal{F}_3$ & $2^{7m}$ \\
         &  $(2^m)^3$ & $\langle u^2\cdot(\alpha_0^{-1}+yh(x))+u y^2\rangle$, & $h(x)\in \mathcal{F}_3$ & $2^{6m}$ \\
         &  $(2^m)^2$ & $\langle u^2\cdot(\alpha_0^{-1}y+y^2h(x))+u y^3\rangle$, & $h(x)\in \mathcal{F}_2$ & $2^{5m}$ \\ \hline
{\bf 2.} &  $(2^m)^2$ & $\langle u^2\cdot y^2h(x)+u^3\rangle$, & $h(x)\in \mathcal{F}_2$ & $2^{4m}$ \\
         &  $2^m$ & $\langle u^2\cdot y^3h+u^3y\rangle$, & $h\in \mathbb{F}_{2^m}$ & $2^{3m}$ \\
         &  $2^m$ & $\langle u^2\cdot y^3h+u^3y^2\rangle$, & $h\in \mathbb{F}_{2^m}$ & $2^{2m}$ \\
         &  $1$ & $\langle u^3y^2\rangle$ & & $2^{m}$ \\ \hline
{\bf 3.} &  $1$ & $\langle 1\rangle$ & & $2^{16m}$ \\
         &  $3$ & $\langle y^s\rangle$, & $1\leq s\leq 3$ & $2^{m(16-2s)}$ \\
         &  $4$ & $\langle u^2y^l\rangle$, & $0\leq l\leq 3$ & $2^{m(8-2l)}$ \\
         &  $1$ & $\langle 0\rangle$ & & $1$ \\
{\bf 4.} &  $1$ & $\langle u, y\rangle$ & & $2^{15m}$ \\ \hline
         &  $2$ & $\langle uy^s, y^{s+1}\rangle$, & $s=1,2$ & $2^{m(15-2s)}$ \\
         &  $1$ & $\langle uy^2, u^2\rangle$ & & $2^{9m}$ \\
         &  $3$ & $\langle u^3y^l, u^2y^{l+1}\rangle$ & $0\leq l\leq 2$ & $2^{m(7-2l)}$
\\ \hline
{\bf 5.} &  $2\cdot 2^m$ & $\langle y^{s+1}h+uy^s, y^{s+2}\rangle$, & $h\in \mathbb{F}_{2^m}$, $s=0,1$ & $2^{m(14-2s)}$ \\
         &  $2^m$ & $\langle y^{3}h+uy^2, u^{2}\rangle$, & $h\in \mathbb{F}_{2^m}$ & $2^{10m}$ \\
         &  $2^m$ & $\langle u^{2}h+uy^3, u^{2}y\rangle$, & $h\in \mathbb{F}_{2^m}$ & $2^{8m}$ \\
         &  $2\cdot 2^m$ & $\langle u^2y^{l+1}h+u^3y^l, u^2y^{l+2}\rangle$, & $h\in \mathbb{F}_{2^m}$, $l=0,1$ & $2^{m(6-2l)}$ \\
         &  $2^m$ & $\langle y^{2}h+u, y^{3}\rangle$, & $h\in \mathbb{F}_{2^m}$ & $2^{13m}$ \\
         &  $2^m$ & $\langle y^{3}h+uy, u^{2}\rangle$, & $h\in \mathbb{F}_{2^m}$ & $2^{11m}$ \\
         &  $2\cdot 2^m$ & $\langle u^2y^{l}h+uy^{l+2}, u^2y^{l+1}\rangle$, & $h\in \mathbb{F}_{2^m}$, $l=0,1$ & $2^{m(9-2l)}$ \\
         &  $2^m$ & $\langle u^2y^{2}h+u^3, u^{2}y^3\rangle$, & $h\in \mathbb{F}_{2^m}$ & $2^{5m}$ \\
         &  $(2^m)^2$ & $\langle y^{2}h(x)+u, u^{2}\rangle$, & $h(x)\in \mathcal{F}_{2}$ & $2^{12m}$ \\
         &  $(2^m)^2$ & $\langle y^{3}h_0+u^2h_1+uy, u^{2}y\rangle$, & $h_0,h_1\in \mathbb{F}_{2^m}$ & $2^{10m}$ \\
         &  $2\cdot(2^m)^2$ & $\langle u^2y^{l}h(x)+uy^{l+2}, u^{2}y^{l+2}\rangle$, & $h(x)\in \mathcal{F}_{2}$,
            $l=0,1$ & $2^{m(8-2l)}$ \\ \hline
{\bf 6.} &  $(2^m)^2$ & $\langle \alpha_0y^2+y^3h(x)+u, u^{2}y\rangle$, & $h(x)\in \mathcal{F}_{2}$ & $2^{11m}$ \\
         &  $(2^m)^2$ & $\langle \alpha_0y^3+u^2h(x)+uy, u^{2}y^2\rangle$, & $h(x)\in \mathcal{F}_{2}$ & $2^{9m}$ \\
         &  $(2^m)^2$ & $\langle u^2(\alpha_0^{-1}+yh(x))+uy^2, u^{2}y^3\rangle$, & $h(x)\in \mathcal{F}_{2}$ & $2^{7m}$ \\
         &  $(2^m)^3$ & $\langle \alpha_0y^2+y^3h(x)+u, u^{2}y^2\rangle$, &  &  \\
                  &   & $y^3h(x)=y^3h_0+u^2h_1+u^2yh_2$ & $h_0,h_1,h_2\in \mathbb{F}_{2^m}$ & $2^{10m}$ \\
         & $(2^m)^3$ & $\langle \alpha_0y^3+u^2h(x)+uy, u^{2}y^3\rangle$, & $h(x)\in \mathcal{F}_{3}$ & $2^{8m}$  \\
         & $(2^m)^3$ & $\langle \alpha_0y^2+u^2h(x)+u, u^{2}y^3\rangle$, & $h(x)\in \mathcal{F}_{3}$ & $2^{9m}$ \\
 \hline
\end{tabular}
\end{center}  }
\noindent
\textit{where $\mathcal{L}$ is the number of cyclic codes in the same row}.

\vskip 3mm \noindent
  {\bf Remark} When $m=1$, the number of $(\delta+\alpha u^2)$-constacyclic codes
of length $4$ over $\frac{\mathbb{F}_{2}[u]}{\langle u^4\rangle}$ is equal to $N_{(2,1,2^{3})}=135$.
But in Example 1 of [23], the authors
list all distinct $(1+u^2)$-constacyclic codes of length $4$ over $\frac{\mathbb{F}_{2}[u]}{\langle u^4\rangle}$
as $131$ distinct $(1+u^2)$-constacyclic codes in  Pages 3119--3120 of [23].

\vskip 3mm \par
  Finally, let $\delta=1$. Then one can easily verify the following
conclusion for $(1+\alpha u^2)$-constacyclic
codes of length $4$. Here we omit the process of proof.

\vskip 3mm \noindent
  {\bf Theorem 3.4} \textit{All distinct self-dual $(1+\alpha u^2)$-constacyclic
codes over $\frac{\mathbb{F}_{2^m}[u]}{\langle u^4\rangle}$ of length $4$ are given by the following four cases}:

\begin{description}
\item{(i)}
  $C=\langle u^2\rangle$.

\item{(ii)}
  \textit{$C=\langle u^2b_0+u(x+1)^3,u^2(x+1)\rangle$, where $b_0\in \mathbb{F}_{2^m}$}.

\item{(iii)}
  \textit{$C=\langle u^2h(x)+u(x+1)^2,u^2(x+1)^2\rangle$, where $h(x)=b_1+b_2(x+1)$ and $b_1,b_2\in \mathbb{F}_{2^m}$}.

\item{(iii)}
  \textit{$C=\langle \alpha_0(x+1)^3+u^2h(x)+u(x+1),u^2(x+1)^3\rangle$, where $h(x)=\alpha_0+b_3(x+1)+b_4(x+1)^2$ and $b_3,b_4\in \mathbb{F}_{2^m}$}.
\end{description}

\par
 \textit{Therefore, the number of self-dual $(1+\alpha u^2)$-constacyclic
codes over $\frac{\mathbb{F}_{2^m}[u]}{\langle u^4\rangle}$ of length $4$ is $1+2^m+2(2^m)^2$}.

\vskip 3mm \par
  Finally, let $m=1$. By Theorem 3.4 we deduce that
there are $11$ self-dual $(1+u^2)$-constacyclic
codes of length $4$ over $\mathbb{F}_2[u]/\langle u^4\rangle$ given by:
\begin{description}
\item{(i)}
  $\langle u^2\rangle$.

\item{(ii)}
  $\langle u(x+1)^3,u^2(x+1)\rangle$, $\langle u^2+u(x+1)^3,u^2(x+1)\rangle$.

\item{(iii)}
  $\langle u^2+u(x+1)^2,u^2(x+1)^2\rangle$,
  $\langle u^2+u(x+1)^2,u^2(x+1)^2\rangle$, \\
  $\langle u^2(x+1)+u(x+1)^2,u^2(x+1)^2\rangle$,
  $\langle u^2x+u(x+1)^2,u^2(x+1)^2\rangle$.

\item{(iii)}
  $\langle (x+1)^3+u^2+u(x+1),u^2(x+1)^3\rangle$, \\
  $\langle (x+1)^3+u^2x+u(x+1),u^2(x+1)^3\rangle$, \\
  $\langle (x+1)^3+u^2x^2+u(x+1),u^2(x+1)^3\rangle$, \\
  $\langle (x+1)^3+u^2(1+x+x^2)+u(x+1),u^2(x+1)^3\rangle$.
\end{description}


\section{Proof of Theorem 2.5} \label{}
\noindent
   In this section, we give a detailed proof of Theorem 2.5.

\par
  Let $\mathcal{C}$ be a $(\delta+\alpha u^2)$-constacyclic codes
over $R$ of length $2^kn$, here we regard $\mathcal{C}$ as an ideal
of the ring $\mathcal{A}+\mathcal{A}$ under the ring isomorphism $\Psi$
determined by Equations (1) and (2) in Section 2. Then by Lemma 2.2(i) and Lemma 2.4(ii),
for each integer $j$, $1\leq j\leq r$, there is a unique ideal $C_j$
of the ring $\mathcal{K}_j+u\mathcal{K}_j$ such that
\begin{equation}
\mathcal{C}=\bigoplus_{j=1}^r\varepsilon_j(x)C_j=\sum_{j=1}^r\varepsilon_j(x)C_j
\ ({\rm mod} \ (x^n+\delta_0)^{2^kn}).
\end{equation}
Moreover, the number of codewords in $\mathcal{C}$ equals $|\mathcal{C}|=\prod_{j=1}^r|C_j|$.

\par
  Now, let $1\leq j\leq r$ and denote $\pi_j=f_j(x)\in \mathcal{K}_j$. We consider how to determine all ideals of the ring $\mathcal{K}_j+u\mathcal{K}_j$
($u^2=\omega_j^2\pi_j^{2^k}$). Since $\mathcal{K}_j$ is a subring of
$\mathcal{K}_j+u\mathcal{K}_j$, we see that $\mathcal{K}_j+u\mathcal{K}_j$ is a free $\mathcal{K}_j$-module
of rank $2$ with the basis $\{1,u\}$. Now, we define
$$\theta: \mathcal{K}_j^2\rightarrow \mathcal{K}_j+u\mathcal{K}_j
\ {\rm via} \ (a_0,a_1)\mapsto a_0+ua_1 \ (\forall a_0,a_1\in \mathcal{K}_j).$$
One can easily verify that $\theta$ is an $\mathcal{K}_j$-module isomorphism from $\mathcal{K}_j^2$
onto $\mathcal{K}_j+u\mathcal{K}_j$. The following lemma can be verified by an argument similar to the proof of
Lemma 3.7 of [7]. Here, we omit its proof.

\vskip 3mm
\noindent
  {\bf Lemma 4.1} \textit{Using the notations above, $C$ is an ideal
of the ring $\mathcal{K}_j+u\mathcal{K}_j$ $(u^2=\omega_j^2\pi_j^{2^k})$ if and only if
there is a unique $\mathcal{K}_j$-submodule $S$ of $\mathcal{K}_j^2$ satisfying}
\begin{equation}
(\omega_j^2\pi_j^{2^k}a_1,a_0)\in S, \ \forall (a_0,a_1)\in S
\end{equation}
\textit{such that $C=\theta(S)$}.

\vskip 3mm \par
  Recall that every $\mathcal{K}_j$-submodule of $\mathcal{K}_j^2$ is called a \textit{linear code}
 over the finite chain ring $\mathcal{K}_j$ of length $2$. A general discussion and description for
linear codes over arbitrary finite chain ring can be found in [24].
 Let $S$ be a linear code over $\mathcal{K}_j$ of length $2$. A matrix $G$ is called a \textit{generator matrix} for $S$ if every codeword in $S$
is a $\mathcal{K}_j$-linear combination of the row vectors of $G$ and
any row vector of $G$ can not be written as a $\mathcal{K}_j$-linear combination of the other row vectors of $G$.

\par
  In the following lemma, we use lowercase letters to denote the elements of $\mathcal{K}_j$ and
$\mathcal{K}_j/\langle \pi_j^l\rangle$ ($1\leq l\leq 2^k\lambda-1$)
in order to simplify the expressions.

\vskip 3mm \noindent
   {\bf Lemma 4.2}  (cf. [8] Lemma 2.2 and Example 2.5) \textit{Using the notations above,
$\sum_{i=0}^{2^k\lambda}(2i+1)2^{md_j(2^k\lambda-i)}$ is the number of
linear codes over $\mathcal{K}_j$ of length $2$}.

\par
   \textit{Moreover, every linear code over
$\mathcal{K}_j$ of length $2$ has one and only one of the following matrices $G$ as their generator matrices}:

\vskip 2mm \par
(i) \textit{$G=(1,a)$, $a\in \mathcal{K}_j$}.

\vskip 2mm \par
(ii) \textit{$G=(\pi_j^s,\pi_j^{s}a)$, $a\in \mathcal{K}_j/\langle \pi_j^{2^k\lambda-s}\rangle$, $1\leq s\leq 2^k\lambda-1$}.

\vskip 2mm \par
(iii) \textit{$G=(\pi_j b,1)$, $b\in \mathcal{K}_j/\langle \pi_j^{2^k\lambda-1}\rangle$}.

\vskip 2mm \par
(iv) \textit{$G=(\pi_j^{s+1}b,\pi_j^s)$, $b\in \mathcal{K}_j/\langle \pi_j^{2^k\lambda-1-s}\rangle$, $1\leq s\leq 2^k\lambda-1$}.

\vskip 2mm \par
  (v) \textit{$G=\left(\begin{array}{cc}\pi_j^s & 0\cr0 & \pi_j^s\end{array}\right)$, $0\leq s\leq 2^k\lambda$}.

\vskip 2mm \par
  (vi) \textit{$G=\left(\begin{array}{cc}1 & c\cr
0 &\pi_j^t\end{array}\right)$,  $c\in \mathcal{K}_j/\langle \pi_j^{t}\rangle$, $1\leq t\leq 2^k\lambda-1$}.

\vskip 2mm \par
  (vii) \textit{$G=\left(\begin{array}{cc}\pi_j^s & \pi_j^sc\cr
0 &\pi_j^{s+t}\end{array}\right)$,  $c\in \mathcal{K}_j/\langle \pi_j^{t}\rangle$, $1\leq t\leq 2^k\lambda-1-s$, $1\leq s\leq 2^k\lambda-2$}.

\vskip 2mm \par
    (viii) \textit{$G=\left(\begin{array}{cc}c & 1\cr \pi_j^t & 0\end{array}\right)$, $c\in \pi_j(\mathcal{K}_j/\langle \pi_j^{t}\rangle)$, $1\leq t\leq 2^k\lambda-1$}.

\vskip 2mm \par
    (ix) \textit{$G=\left(\begin{array}{cc}\pi_j^sc & \pi_j^s\cr \pi_j^{s+t} & 0\end{array}\right)$, $c\in \pi_j(\mathcal{K}_j/\langle \pi_j^{t}\rangle)$,
$1\leq t\leq 2^k\lambda-1-s$, $1\leq s\leq 2^k\lambda-2$}.

\vskip 3mm\par
   Let $\beta\in \mathcal{K}_j$ and $\beta\neq 0$. By Lemma 2.3(ii), there is a unique integer $t$, $0\leq t\leq 2^k\lambda-1$,
such that $\beta=\pi_j^tw$ for some $w\in\mathcal{K}_j^\times$.
We call $t$ the \textit{$\pi_j$-degree} of $\beta$ and denote it by $\|\beta\|_{\pi_j}=t$. If $\beta=0$,
we write $\|\beta\|_{\pi_j}=2^k\lambda$.  For any vector $(\beta_1,\beta_2)\in \mathcal{K}_j^2$, we define the \textit{$\pi_j$-degree} of $(\beta_1,\beta_2)$
by $$\|(\beta_1,\beta_2)\|_{\pi_j}={\rm min}\{\|\beta_1\|_{\pi_j},\|\beta_2\|_{\pi_j}\}.$$
Now, as a special case of [24] Proposition 3.2 and Theorem 3.5,
we deduce the following lemma.

\vskip 3mm \noindent
   {\bf Lemma 4.3} (cf. [8] Lemma 2.3) \textit{Let $S$ be a nonzero linear code over $\mathcal{K}_j$ of length $2$, and $G$ be a generator matrix of $S$ with row vectors $G_1,\ldots,G_\rho\in \mathcal{K}_j^2\setminus\{0\}$ satisfying
$$\|G_j\|_{\pi_j}=t_i, \ {\rm where} \ 0\leq t_1\leq\ldots\leq t_\rho\leq 2^k\lambda-1.$$
Then the number of codewords in $S$ is equal to $|S|=|\mathcal{F}_j|^{\sum_{i=1}^\rho(2^k\lambda-t_i)}
=2^{md_j\sum_{i=1}^\rho(2^k\lambda-t_i)}$}.

\vskip 3mm\par
   For any positive integer $i$, let $\lceil\frac{i}{2}\rceil={\rm min}\{l\in\mathbb{Z}^{+}\mid
l\geq \frac{i}{2}\}$ and $\lfloor\frac{i}{2}\rfloor={\rm max}\{l\in\mathbb{Z}^{+}\cup\{0\}\mid
l\leq \frac{i}{2}\}$. It is well known that $\lceil\frac{i}{2}\rceil+\lfloor\frac{i}{2}\rfloor=i$. Using these notations,
we list all distinct $\mathcal{K}_j$-submodules of $\mathcal{K}_j^2$ satisfying Condition (4) in Lemma 4.1
by the following lemma.

\vskip 3mm \noindent
   {\bf Lemma 4.4} \textit{Using the notations in Section 2, every linear code $S$ over
$\mathcal{K}_j$ of length $2$ satisfying Condition $(4)$ in Lemma 4.1
has one and only one of the following matrices $G$ as its generator matrix}:

\begin{description}
   \item{(I)}
   $\sum_{s=0}^{2^{k}\lambda-1}2^{md_j(2^{k-1}\lambda-\lceil\frac{s}{2}\rceil)}$ \textit{matrices given by the following two cases}:

\begin{enumerate}
\vskip 2mm
 \item[(I-1)]
   \textit{$G=(\omega_jf_j(x)^{2^{k-1}+s}
+f_j(x)^{2^{k-1}\lambda+\lceil \frac{s}{2}\rceil}h(x),f_j(x)^s)$ \\
where $h(x)\in  \frac{\mathbb{F}_{2^m}[x]}{\langle f_j(x)^{2^{k-1}\lambda-\lceil \frac{s}{2}\rceil}\rangle}$, if $0\leq s\leq 2^{k-1}\lambda-1$}.

\vskip 2mm
\item[(I-2)]
   \textit{$G=(f_j(x)^{2^{k-1}\lambda+\lceil \frac{s}{2}\rceil}h(x),f_j(x)^s)$ \\
   where $h(x)\in  \frac{\mathbb{F}_{2^m}[x]}{\langle f_j(x)^{2^{k-1}\lambda-\lceil \frac{s}{2}\rceil}\rangle}$,
   if $2^{k-1}\lambda\leq s\leq 2^{k}\lambda-1$}.
\end{enumerate}

\item{(II)}
 $2^{k}\lambda+1$ \textit{matrices}:\\
\begin{enumerate}
 \item[]
  \textit{$G=\left(\begin{array}{cc}f_j(x)^s & 0\cr0 & f_j(x)^s\end{array}\right)$, $0\leq s\leq 2^{k}\lambda$}.
\end{enumerate}

\item{(III)}
 $\sum_{t=1}^{2^{k}\lambda-1}(2^{k}\lambda-t)2^{md_j\lfloor\frac{t}{2}\rfloor}$
 \textit{matrices given by the following three cases}:

\begin{enumerate}
 \item[(III-1)]
   \textit{$G=\left(\begin{array}{cc} 0 & f_j(x)^s\cr f_j(x)^{s+1} & 0\end{array}\right),
\ 0\leq s\leq 2^{k}\lambda-2.$}

\vskip 2mm
\item[(III-2)]
   $G=\left(\begin{array}{cc} f_j(x)^{s+\lceil \frac{t}{2}\rceil}h(x) & f_j(x)^s\cr f_j(x)^{s+t} & 0\end{array}\right)$
   \textit{where $h(x)\in\frac{\mathbb{F}_{2^m}[x]}{\langle f_j(x)^{\lfloor \frac{t}{2}\rfloor}\rangle}$ and
   $0\leq s\leq 2^{k}\lambda-1-t$, if $2\leq t\leq 2^k$}.

\vskip 2mm
\item[(III-3)]
   $G=\left(\begin{array}{cc} \omega_jf_j(x)^{2^{k-1}+s}+f_j(x)^{s+\lceil \frac{t}{2}\rceil}h(x) & f_j(x)^s
   \cr f_j(x)^{s+t} & 0\end{array}\right)$
   \textit{where $h(x)\in\frac{\mathbb{F}_{2^m}[x]}{\langle f_j(x)^{\lfloor \frac{t}{2}\rfloor}\rangle}$ and
   $0\leq s\leq 2^{k}\lambda-1-t$, if $2^k+1\leq t\leq 2^{k}\lambda-1$}.
\end{enumerate}
\end{description}

\noindent
  {\bf Proof.} In order to simplify the symbol, we write
$\mathcal{K}_j$, $\mathcal{F}_j$, $\pi_j=f_j(x)$ and $\omega_j$ simply as $\mathcal{K}$, $\mathcal{F}$, $\pi=f(x)$ and $\omega$ respectively, in the following.
  Let $S$ be the $\mathcal{K}$-submodule
of $\mathcal{K}^2$ with generator matrix $G$, where $G$ is a matrix given by Lemma 4.2.  Then we only need to consider the following five cases.

\par
  {\bf Case 1}. Let $G$ be given by Lemma 4.2 (i) and (ii), i.e., $G=(\pi^s,\pi^sa)$ where $a\in \mathcal{K}/\langle \pi^{2^{k}\lambda-s}\rangle$
and $0\leq s\leq 2^{k}\lambda-1$.

   Suppose that $S$ satisfies Condition $(4)$ in Lemma 4.1. Then
$$(\omega^2\pi^{2^k}\cdot \pi^sa, \pi^s)=b(\pi^s,\pi^sa)=(\pi^sb,\pi^sab)
\ {\rm for} \ {\rm some} \ b\in \mathcal{K},$$
which is equivalent to that $\omega^2\pi^{2^k+s}a=\pi^sb$ and $\pi^s=\pi^sab$. The latter implies
$$\pi^s=(\pi^sb)a=\omega^2\pi^{2^k+s}a^2=\pi^{2^k+s}\cdot \omega^2a^2 \ {\rm in} \ \mathcal{K}.$$
From this and by $0\leq s<{\rm min}\{2^{k}\lambda,2^k+s\}$, we obtain a contradiction. Hence
$S$ does not satisfy Condition (4) in Lemma 4.1.

\par
  {\bf Case 2}. Let $G$ be given by Lemma 4.2 (iii) and (iv), i.e., $G=(\pi^{s+1}b,\pi^s)$ where $b\in \mathcal{K}/\langle \pi^{2^{k}\lambda-1-s}\rangle$ and $0\leq s\leq 2^{k}\lambda-1$.

  Then
$S$ satisfies Condition (4) in Lemma 4.1 if and only if there exists $a\in \mathcal{K}$
such that
$(\omega^2\pi^{2^k}\cdot \pi^s, \pi^{s+1}b)=a(\pi^{s+1}b,\pi^s)=(\pi^{s+1}ab,\pi^sa)$, i.e.,
$\omega^2\pi^{2^k+s}=\pi^{s+1}ab$ and $\pi^{s+1}b=\pi^sa$. These conditions are simplified to
that $b\in \mathcal{K}/\langle \pi^{2^{k}\lambda-1-s}\rangle$ satisfying
$\omega^2\pi^{2^k+s}=(\pi^sa)\pi b=\pi^{s+2}b^2$. As $\omega$ is an invertible
element of $\mathcal{K}$ by Lemma 2.4(i), we can set $z=\omega^{-1}b\in \mathcal{K}/\langle \pi^{2^{k}\lambda-1-s}\rangle$.
Then the Condition (4) is equivalent to
\begin{equation}
\pi^{2^k+s}=\pi^{s+2}z^2 \ {\rm and} \ b=\omega z,
\ {\rm where} \ z\in \mathcal{K}/\langle \pi^{2^{k}\lambda-1-s}\rangle.
\end{equation}
Then we have one of the following subcases:

\par
  {\bf Subcase 2.1}. If $0\leq s\leq 2^k(\lambda-1)-1$, we have $2^k\leq 2^k+s\leq 2^{k}\lambda-1$.
Now, by Lemma 2.3(ii) we may assume $z=\sum_{i=0}^{2^{k}\lambda-2-s}z_i\pi^i\in \mathcal{K}/\langle \pi^{2^{k}\lambda-1-s}\rangle$
where $z_i\in \mathcal{F}$ for all $i=0,1,\ldots, 2^{k}\lambda-2-s$. In the the ring $\mathcal{K}$ we have
$$\pi^{s+2}z^2=\sum_{i=0}^{2^{k}\lambda-2-s}z_i^2\pi^{s+2+2i} \ {\rm in} \ {\rm which} \ \pi^{2^{k}\lambda}=0.$$
Hence $b=\omega z$ satisfies Condition (5) if and only if the elements $z_i\in \mathcal{F}$ satisfying:
\begin{enumerate}
   \item[$\diamond$]
   $z_0^2+z_1^2\pi^2+\ldots+z_{2^{k-1}-2}^2\pi^{2^k-4}\equiv0$ (mod $\pi^{2^k-2}$), i.e. $(z_0+z_1\pi+\ldots+z_{2^{k-1}-1}\pi^{2^{k-1}-2})^2\equiv 0$ (mod $\pi^{2^k-2}$)
   in $\mathcal{K}$.
   The letter is equivalent to that \\
   $z_i=0$ for all integer $i$: $0\leq i\leq 2^{k-1}-2$.

\item[$\diamond$]
   $z_{2^{k-1}-1}=1$.

\item[$\diamond$]
 $z_i=0$, if $2^k+s<s+2+2i<2^{k}\lambda$, i.e., $2^{k-1}\leq i\leq 2^{k-1}\lambda-2-\lfloor \frac{s}{2}\rfloor$.
\end{enumerate}

\noindent
  As stated above, we conclude that
$$b=\omega \pi^{2^{k-1}-1}+\sum_{2^{k-1}\lambda-1-\lfloor\frac{s}{2}\rfloor\leq i\leq 2^{k}\lambda-2-s}b_i\pi^i,
\ {\rm where} \ b_i\in \mathcal{F} \ {\rm arbitrary}.$$
By $\lceil\frac{s}{2}\rceil+\lfloor\frac{s}{2}\rfloor=s$, we have
$(s+1)+(2^{k-1}\lambda-1-\lfloor\frac{s}{2}\rfloor)=2^{k-1}\lambda+\lceil \frac{s}{2}\rceil$ and
$$(2^{k}\lambda-2-s)-(2^{k-1}2^{k-1}-1-1-\lfloor\frac{s}{2}\rfloor)=2^{k-1}\lambda-\lceil \frac{s}{2}\rceil-1.$$
From these, we deduce
\begin{eqnarray*}
\pi^{s+1}b &=& \omega \pi^{2^{k-1}+s}
+\pi^{2^{k-1}\lambda+\lceil \frac{s}{2}\rceil}\sum_{2^{k-1}\lambda-1-\lfloor\frac{s}{2}\rfloor\leq i\leq 2^{k}\lambda-2-s}
b_iy^{i-(2^{k-1}\lambda-1-\lfloor\frac{s}{2}\rfloor)}\\
&=&\omega \pi^{2^{k-1}+s}
+\pi^{2^{k-1}\lambda+\lceil \frac{s}{2}\rceil}h(x),
\end{eqnarray*}
where $h(x)=\sum_{i=0}^{2^{k-1}\lambda-\lceil \frac{s}{2}\rceil-1}h_iy^i=\sum_{i=0}^{2^{k-1}\lambda-\lceil \frac{s}{2}\rceil-1}h_if(x)^i\in \frac{\mathbb{F}_{2^m}[x]}{\langle f(x)^{2^{k-1}\lambda-\lceil \frac{s}{2}\rceil}\rangle}$ with $h_i\in \mathcal{F}$ for all $i$.

\par
  {\bf Subcase 2.2}. If $2^k(\lambda-1)\leq s\leq 2^{k}\lambda-1$, we have $\pi^{2^k+s}=0$ and that
the condition (5) is simplified to $\pi^{s+2}b^2=0$. The latter condition is equivalent
to $s+2+2\|b\|_{\pi}\geq 2^{k}\lambda$, i.e., $\|b\|_{\pi}\geq 2^{k-1}\lambda-1-\lfloor \frac{s}{2}\rfloor$,
and hence
$$b\in \pi^{2^{k-1}\lambda-1-\lfloor \frac{s}{2}\rfloor}\left(\mathcal{K}/\langle \pi^{2^{k}\lambda-1-s}\rangle\right)=f(x)^{2^{k-1}\lambda-1-\lfloor \frac{s}{2}\rfloor}\left(\mathcal{K}/\langle f(x)^{2^{k}\lambda-1-s}\rangle\right).$$
In this case,
by $(2^{k}\lambda-1-s)-(2^{k-1}\lambda-1-\lfloor \frac{s}{2}\rfloor-1)=2^{k-1}\lambda-\lceil \frac{s}{2}\rceil$ we have
$$\pi^{s+1}b=f(x)^{2^{k-1}\lambda+\lceil \frac{s}{2}\rceil}h(x), \ {\rm where}
\ h(x)\in  \frac{\mathbb{F}_{2^m}[x]}{\langle f(x)^{2^{k-1}\lambda-\lceil \frac{s}{2}\rceil}\rangle}.$$

\par
  Now, the conclusions in (I) follows from
$|\frac{\mathbb{F}_{2^m}[x]}{\langle f(x)^{2^{k-1}\lambda-\lceil \frac{s}{2}\rceil}\rangle}|=2^{md_j(2^{k-1}\lambda-\lceil \frac{s}{2}\rceil)}$.

\par
   {\bf Case 3}. Let $G$ be given by Lemma 4.2(v). Then it is obvious that every submodule $S$ satisfies Condition (4) in Lemma 4.1.

\par
   {\bf Case 4}. Let $S$ be generated by matrix $G$ where $G$ is given by Lemma 4.2 (vi) and (vii), i.e.,
$G=\left(\begin{array}{cc}\pi^s & \pi^sc\cr
0 &\pi^{s+t}\end{array}\right)$,  $c\in \mathcal{K}/\langle \pi^{t}\rangle$, $1\leq t\leq 2^{k}\lambda-1-s$ and $0\leq s\leq 2^{k}\lambda-2$.
Suppose that $S$ satisfies Condition (4) in Lemma 4.1. Then $(\omega^2\pi^{2^k}\cdot \pi^sc, \pi^s)\in S$. Hence
there exist $a,b\in\mathcal{K}$ such that
$$(\omega^2\pi^{2^k}\cdot \pi^sc, \pi^s)=a(\pi^s,\pi^sc)+b(0,\pi^{s+t})=(\pi^sa, \pi^sac+\pi^{s+t}b),$$
which is equivalent to $\omega^2\pi^{2^k+s}c=\pi^sa$ and $\pi^s=\pi^sac+\pi^{s+t}b$. These imply
$$\pi^s=\pi^{s+1}\left(\omega^2\pi^{2^k-1+s}c^2+\pi^{s+t-1}b\right),$$
and we get a contradiction.

\par
   {\bf Case 5}. Let $G$ be given by Lemma 4.2 (viii) and (ix). Then we have $G=\left(\begin{array}{cc} \pi^sc & \pi^s\cr \pi^{s+t} & 0\end{array}\right)$, $c\in \pi(\mathcal{K}/\langle \pi^{t}\rangle)$,
$1\leq t\leq 2^{k}\lambda-1-s$ and $0\leq s\leq 2^{k}\lambda-2$. It is clear that
the conditions $1\leq t\leq 2^{k}\lambda-1-s$ and $0\leq s\leq 2^{k}\lambda-2$ is equivalent to
\begin{equation}
0\leq s\leq 2^{k}\lambda-1-t \ {\rm and}
\ 1\leq t\leq 2^{k}\lambda-1.
\end{equation}
  Obviously, we have $(\omega^2\pi^{2^k}\cdot 0, \pi^{s+t})=\pi^t(\pi^sc, \pi^s)-c(\pi^{s+t},0)\in S$. Hence
$S$ satisfies Condition (4) if and only if there exist
$a,b\in\mathcal{K}$ such that
$$(\omega^2\pi^{2^k}\cdot \pi^s, \pi^sc)=a(\pi^sc,\pi^s)+b(\pi^{s+t},0)=(\pi^sac+\pi^{s+t}b, \pi^sa),$$
which is equivalent to $\omega^2\pi^{2^k+s}=\pi^sac+\pi^{s+t}b$ and $\pi^sc=\pi^sa$. These conditions are simplified
to
\begin{equation}
\omega^2\pi^{2^k+s}=\pi^sc^2+\pi^{s+t}b \ {\rm for} \ {\rm some} \ b\in \mathcal{K}.
\end{equation}
Then we have one of the following four subcases:

\par
  {\bf Subcase 5.1}.
  If $t=1$, we have $\pi(\mathcal{A}/\langle \pi^{t}\rangle)=\{0\}$. In this case, it is obvious that
$c=0$ satisfies $\omega^2\pi^{2^k+s}=\pi^s\cdot 0^2+\pi^{s+t}b$ where $b=\omega^2\pi^{2^k-t}$. Then we have
$2^{k+1}-1$ matrices satisfing Condition (1) in Lemma 2.2:
$$G=\left(\begin{array}{cc} 0 & \pi^s\cr \pi^{s+1} & 0\end{array}\right),
\ 0\leq s\leq 2^{k}\lambda-2.$$

\par
  {\bf Subcase 5.2}.
  If $t=2$, we have $\pi(\mathcal{K}/\langle \pi^{t}\rangle)=\{c_1\pi\mid c_1\in \mathcal{F}\}$. In this case,
$$\omega^2\pi^{2^k+s}=\pi^{s+2}c_1^2+\pi^{s+2}b
\ {\rm where} \ b=c_1^2+\omega^2\pi^{2^k-2},
\ \forall c_1\in \mathcal{F}.$$
 Then we have
$|\mathcal{F}|=2^{md_j}$ matrices:
$G=\left(\begin{array}{cc} \pi^{s+1}c_1 & \pi^s\cr \pi^{s+2} & 0\end{array}\right)$,
where $c_1\in \mathcal{F}$ arbitrary.

\par
  In the following we assume $t\geq 3$. Set $g=\omega^{-1}c\in \pi(\mathcal{K}/\langle \pi^{t}\rangle)$.
Then $c=\omega g$ and Equation (7) is equivalent to
\begin{equation}
\pi^{2^k+s}=\pi^sg^2+\pi^{s+t}b \ {\rm for} \ {\rm some} \ b\in \mathcal{K}.
\end{equation}
Now, let $g=\sum_{i=1}^{t-1}g_i\pi^i\in \pi(\mathcal{K}/\langle \pi^{t}\rangle)$
where $g_i\in \mathcal{F}$ for all $i=1,\ldots, t-1$,
and $b=\sum_{j=0}^{2^{k}\lambda-1}b_j\pi^j$ where $b_j\in \mathcal{F}$ for all $j$.
Then in the the ring $\mathcal{K}$, we have
$$\pi^{s}g^2+\pi^{s+t}b=\sum_{i=1}^{t-1}g_i^2\pi^{s+2i}+\sum_{j=0}^{2^{k}\lambda-1-(s+t)}b_j\pi^{s+t+j} \ {\rm in} \ {\rm which} \ \pi^{2^{k}\lambda}=0.$$

\par
  {\bf Subcase 5.3}.
  Let $t=2l+1\leq 2^{k}\lambda-1$ where $1\leq l\leq 2^{k-1}\lambda-1$. Then $l=\lfloor \frac{t}{2}\rfloor$ and
$l+1=\lceil \frac{t}{2}\rceil$.
We have two situations:
\begin{enumerate}
 \item[($\dag$)]
  When $1\leq l\leq 2^{k-1}-1$, we have $3\leq t\leq 2^{k}-1$. In this case, the element
  $g\in \pi(\mathcal{K}/\langle \pi^{t}\rangle)$ satisfies Condition (8) if and only if
$$g_i=0 \ {\rm for} \ {\rm all} \ 1\leq i\leq l,$$
i.e. $g=\sum_{i=l+1}^{t-1}g_i\pi^i$, where $g_{l+1},\ldots,g_{t-1}\in \mathcal{F}$ arbitrary.
In fact, we have
$\pi^{2^k+s}=\pi^{s}g^2+\pi^{s+t}b
\ {\rm with} \ b=\pi^{2^k-t}+\sum_{i=l+1}^{t-1}g_i^2\pi^{2i-t}\in \mathcal{K}.$

\par
  As stated above, we have $(t-1)-(l+1)=t-\lceil \frac{t}{2}\rceil-1=\lfloor \frac{t}{2}\rfloor-1$ and
$$c=\omega g=\sum_{i=l+1}^{t-1}c_i\pi^i=\pi^{l+1}\sum_{i=l+1}^{t-1}c_i\pi^{i-(l+1)}=\pi^{\lceil \frac{t}{2}\rceil}h(x),$$
where $h(x)=\sum_{i=l+1}^{t-1}c_i\pi^{i-(l+1)}
=\sum_{j=0}^{\lfloor \frac{t}{2}\rfloor-1}c_{j+\lceil\frac{t}{2}\rceil}f(x)^{j}$.
Hence there are $2^{md_j\lfloor\frac{t}{2}\rfloor}$ matrices satisfying Condition (4) in Lemma 4.1:
\begin{equation}
G=\left(\begin{array}{cc} \pi^{s+\lceil \frac{t}{2}\rceil}h(x) & \pi^s\cr \pi^{s+t} & 0\end{array}\right),
\ h(x)\in\frac{\mathbb{F}_{2^m}[x]}{\langle f(x)^{\lfloor \frac{t}{2}\rfloor}\rangle}.
\end{equation}

 \item[($\ddag$)]
  When $2^{k-1}\leq l\leq 2^{k-1}\lambda-1$, we have $2^k+1\leq t=2l+1\leq 2^{k}\lambda-1$. This implies
  $s+t\geq 2^k+s+1$. In this case, the element $c\in \pi(\mathcal{K}/\langle \pi^{t}\rangle)$ satisfies Condition (7) if and only if
$$c=\omega \pi^{2^{k-1}}+\sum_{i=l+1}^{t-1}c_i\pi^i, \ {\rm where} \ c_{l+1},\ldots,c_{t-1}\in \mathcal{F}
\ {\rm arbitrary}.$$
In fact, we have $\omega^2\pi^{2^k+s}=\pi^sc^2+\pi^{s+t}b$ where
$b=\sum_{i=l+1}^{t-1}c_i^2\pi^{2i-t}\in \mathcal{K}$.
Hence there are $2^{md_j\lfloor\frac{t}{2}\rfloor}$ matrices satisfying Condition (4) in Lemma 4.1:
\begin{equation}
G=\left(\begin{array}{cc} \omega \pi^{2^{k-1}+s}+\pi^{s+\lceil \frac{t}{2}\rceil}h(x) & \pi^s\cr \pi^{s+t} & 0\end{array}\right),
\ h(x)\in\frac{\mathbb{F}_{2^m}[x]}{\langle f(x)^{\lfloor \frac{t}{2}\rfloor}\rangle}.
\end{equation}
\end{enumerate}

\par
  {\bf Subcase 5.4}.
  Let $t=2l\leq 2^{k}\lambda-1$ where $2\leq l\leq 2^{k-1}\lambda-1$. Then $l=\lfloor \frac{t}{2}\rfloor=\lceil \frac{t}{2}\rceil$.
We have two situations:
\begin{enumerate}
 \item[($\dag$)]
  When $2\leq l\leq 2^{k-1}$, we have $4\leq t\leq 2^{k}$. In this case, the element $c\in y(\mathcal{A}/\langle y^{t}\rangle)$ satisfies Condition (7) if and only if
$$c=\sum_{i=l}^{t-1}c_i\pi^i, \ {\rm where} \ c_{l},\ldots,c_{t-1}\in \mathcal{F} \ {\rm arbitrary}.$$
In fact, for any $c_{l},\ldots,c_{t-1}\in \mathcal{F}$ we have
$$\omega^2\pi^{2^k+s}=\pi^{s}c^2+\pi^{s+t}b
\ {\rm with} \ b=\omega^2\pi^{2^k-t}+\sum_{i=l}^{t-1}c_i^2\pi^{2i-t}\in \mathcal{K}.$$
\par
  As stated above, we have $(t-1)-l=l-1=\lfloor \frac{t}{2}\rfloor-1$ and
$$c=\sum_{i=l}^{t-1}c_i\pi^i=\pi^{l}\sum_{i=l}^{t-1}c_i\pi^{i-l}=\pi^{\lceil \frac{t}{2}\rceil}h(x),
\ {\rm where} \ h(x)\in\frac{\mathbb{F}_{2^m}[x]}{\langle f(x)^{\lfloor \frac{t}{2}\rfloor}\rangle}.$$
Hence there are $2^{md_j\lfloor\frac{t}{2}\rfloor}$ matrices satisfying Condition (4) in Lemma 4.1 given by
Equation (9).

 \item[($\ddag$)]
  When $2^{k-1}+1\leq l\leq 2^{k-1}\lambda-1$, we have $2^k+2\leq t=2l\leq 2^{k}\lambda-2$. This implies
  $s+t\geq 2^k+s+2$. In this case, the element $c\in \pi(\mathcal{K}/\langle \pi^{t}\rangle)$ satisfies Condition (7) if and only if
$$c=\omega \pi^{2^{k-1}}+\sum_{i=l}^{t-1}c_i\pi^i, \ {\rm where} \ c_{l},\ldots,c_{t-1}\in \mathcal{F} \ {\rm arbitrary}.$$
In fact, we have $\omega^2\pi^{2^k+s}=\pi^sc^2+\pi^{s+t}b$ where
$b=\sum_{i=l}^{t-1}c_i^2\pi^{2i-t}\in \mathcal{K}$.
Hence there are $2^{md_j\lfloor\frac{t}{2}\rfloor}$ matrices satisfying Condition (4) in Lemma 4.1
given by Equation (10).
\end{enumerate}

\par
  As stated above, the number of matrices satisfying Condition (4) in Case (viii) and
(ix) of Lemma 4.2 is equal to
$$\sum_{t=1}^{2^{k}\lambda-1}\sum_{s=0}^{2^{k}\lambda-1-t}2^{md_j\lfloor\frac{t}{2}\rfloor}
=\sum_{t=1}^{2^{k}\lambda-1}(2^{k}\lambda-t)2^{md_j\lfloor\frac{t}{2}\rfloor}$$
by Equation (6).
\hfill
$\Box$

\vskip 3mm\par
  Finally, we give a proof for Theorem 2.5 as follows.

  Let $\mathcal{C}$ be a
$(\delta+\alpha u^2)$-constacyclic code over $R$ of length $2^kn$. Here we regard
$\mathcal{C}$ as an ideal of the ring $\mathcal{A}+u\mathcal{A}$.
Then by Lemma 2.4(ii) for
each integer $j$, $1\leq j\leq r$, there is a unique ideal
$C_j$ of the ring $\mathcal{K}_j+u\mathcal{K}_j$ such that
$$\mathcal{C}=\bigoplus_{j=1}^r\varepsilon_j(x)C_j=\sum_{j=1}^r\varepsilon_j(x)C_j
\ ({\rm mod} \ (x^n+\delta_0)^{2^k\lambda}).$$
This implies $|\mathcal{C}|=\prod_{j=1}^r|C_j|$. Furthermore,
since $C_j$ is an ideal
of the ring $\mathcal{K}_j+u\mathcal{K}_j$ $(u^2=\omega_j^2f_j(x)^{2^k})$, by Lemma 4.1 there is a unique
$\mathcal{K}_j$-submodule $S$ of $\mathcal{K}_j^2$ satisfying Condition (4) such that $C_j=\theta(S)$. From this and by
Lemma 4.4, we deduce that $S$ has one and only one of the following matrices $G$ as its generator matrix:
\begin{description}
\item{{\bf 1.}}
    $G$ is given by Lemma 4.4(I-1). In this case, by the definition of $\theta$ above Lemma 4.1 we have
\begin{eqnarray*}
&&\theta\left(\omega_jf_j(x)^{2^{k-1}+s}
+f_j(x)^{2^{k-1}\lambda+\lceil \frac{s}{2}\rceil}h(x),f_j(x)^s\right)\\
&=&\omega_jy^{2^{k-1}+s}
+f_j(x)^{2^{k-1}\lambda+\lceil \frac{s}{2}\rceil}h(x)+u f_j(x)^s.
\end{eqnarray*}
This implies
$C_j=\theta(S)=\langle \omega_jy^{2^{k-1}+s}
+f_j(x)^{2^{k-1}\lambda+\lceil \frac{s}{2}\rceil}h(x)+u f_j(x)^s\rangle.$

\par
  As $\|(\omega_jf_j(x)^{2^{k-1}+s}
+f_j(x)^{2^{k-1}\lambda+\lceil \frac{s}{2}\rceil}h(x),f_j(x)^s)\|_{f_j(x)}=s$, by Lemma 4.3 it follows that
$|C_j|=|S|=2^{md_j(2^{k}\lambda-s)}$.

\item{{\bf 2.}}
  $G$ is given by Lemma 4.4(I-2). In this case, an argument
similar to (i-1) shows that
$C_j=\theta(S)=\langle f_j(x)^{2^{k-1}\lambda+\lceil \frac{s}{2}\rceil}h(x)+u f_j(x)^s\rangle$ and $|C_j|=|S|=2^{md_j(2^{k}\lambda-s)}$.

\item{{\bf 3.}}
  $G$ is given by Lemma 4.4(II). In this case, we have
$$C_j=\theta(S)=\langle \theta(f_j(x)^s, 0), \theta(0, f_j(x)^s)\rangle=\langle f_j(x)^s, uf_j(x)^s\rangle=\langle f_j(x)^s\rangle.$$
Then by Lemma 4.3 and $\|(f_j(x)^s, 0)\|_{f_j(x)}=\|(0, f_j(x)^s)\|_{f_j(x)}=s$, we obtain
$|C_j|=|S|=2^{md_j((2^{k}\lambda-s)+(2^{k}\lambda-s))}=2^{md_j(2^{k+1}\lambda-2s)}$.

\item{{\bf 4.}}
  $G$ is given by Lemma 4.4(III-1). In this case,
we have
$$C_j=\theta(S)=\langle \theta(0,f_j(x)^s), \theta(f_j(x)^{s+1}, 0)\rangle=\langle uf_j(x)^s, f_j(x)^{s+1}\rangle.$$
Then by Lemma 2.4, $\|(0,f_j(x)^s)\|_{f_j(x)}=s$ and $\|(f_j(x)^{s+1}, 0)\|_{f_j(x)}=s+1$, we obtain
$|C_j|=|S|=2^{md_j((2^{k}\lambda-s)+(2^{k}\lambda-s-1))}=2^{md_j(2^{k+1}\lambda-2s-1)}$.

\item{{\bf 5.}}
   $G$ is given by Lemma 4.4(III-2). In this case,
an argument
similar to (i) shows that
\begin{eqnarray*}
C_j&=&\theta(S)=\langle \theta(f_j(x)^{s+\lceil \frac{t}{2}\rceil}h(x),f_j(x)^s), \theta(f_j(x)^{s+t}, 0)\rangle \\
 &=&\langle f_j(x)^{s+\lceil \frac{t}{2}\rceil}h(x)+uf_j(x)^s, f_j(x)^{s+t}\rangle.
\end{eqnarray*}
Then by Lemma 4.3, we obtain
$|C_j|=|S|=2^{md_j((2^{k}\lambda-s)+(2^{k}\lambda-s-t))}=2^{md_j(2^{k+1}\lambda-2s-t)}$, since$\|(f_j(x)^{s+\lceil \frac{t}{2}\rceil}h(x),f_j(x)^s)\|_{f_j(x)}=s$ and $\|(f_j(x)^{s+t}$, $0)\|_{f_j(x)}=s+t$.

\item{{\bf 6.}}
  $G$ is given by Lemma 4.4(III-3). Then
an argument
similar to (v) shows that
$$C_j=\theta(S)
=\langle \omega_jf_j(x)^{2^{k-1}+s}+f_j(x)^{s+\lceil \frac{t}{2}\rceil}h(x)+uf_j(x)^s, f_j(x)^{s+t}\rangle.$$
and
$|C_j|=|S|=2^{md_j(2^{k+1}\lambda-2s-t)}$.

\end{description}

\par

\par
  Finally, by Lemmas 4.1 and 4.4 we see that the number of ideals in the ring $\mathcal{K}_j+u\mathcal{K}_j$
$$N_{(2^m,d_j,2^{k}\lambda)}=\sum_{s=0}^{2^{k}\lambda-1}2^{md_j(2^{k-1}\lambda-\lceil\frac{s}{2}\rceil)}+(2^{k}\lambda+1)
+\sum_{t=1}^{2^{k}\lambda-1}(2^{k}\lambda-t)2^{md_j\lfloor\frac{t}{2}\rfloor}.$$
This implies $N_{(2^m,d_j,2^{k}\lambda)}=\sum_{i=0}^{2^{k-1}\lambda}(1+4i)2^{md_j(2^{k-1}\lambda-i)}$.
Therefore, Theorem 2.5 was proved.


\section{Conclusions and further research}\label{}
\noindent
We give an explicit representation and enumeration for all distinct
$(\delta+\alpha u^2)$-constacyclic codes over $\frac{\mathbb{F}_{2^m}[u]}{\langle u^{2\lambda}\rangle}$ of length $2^kn$ for
any integers $k,\lambda\geq 2$ and odd positive integer $n$, provide a clear and exact formula to count the number of this class of constacyclic codes and obtain a clear formula to count the number of codewords in each code from its generators directly.

\par
  Our further interest
is to consider the construction of self-dual $(1+\alpha u^2)$-constacyclic codes over $\frac{\mathbb{F}_{2^m}[u]}{\langle u^{4}\rangle}$ of arbitrary even length and study the properties of self-dual codes over $\mathbb{F}_{2^m}$ derived from these codes.



\vskip 3mm \noindent {\bf Acknowledgments}
 Part of this work was done when Yonglin Cao was visiting Chern Institute of Mathematics,
Nankai University, Tianjin, China. Yonglin Cao would like to thank the institution for the kind hospitality. This research is
supported in part by the National Natural Science Foundation of
China (Grant Nos. 11671235, 11801324),  the Shandong Provincial Natural Science Foundation, China
(Grant No. ZR2018BA007), the Scientific Research Fund of Hubei Provincial Key Laboratory of Applied Mathematics (Hubei University)
(Grant No. AM201804) and the Scientific Research Fund of Hunan
Provincial Key Laboratory of Mathematical Modeling and Analysis in
Engineering (No. 2018MMAEZD09). H.Q. Dinh is
grateful for the Centre of Excellence in Econometrics, Chiang Mai University,
Thailand, for partial financial support.




\begin{thebibliography}{24}

\bibitem{s1} T. Abualrub, I. Siap, Constacyclic codes over $\mathbb{F}_2+u\mathbb{F}_2$,
J. Franklin Inst. {\bf 346} (2009), 520--529.

\bibitem{s2} M. C. V. Amerra, F. R. Nemenzo, On $(1-u)$-cyclic codes over $\mathbb{F}_{p^k}+u\mathbb{F}_{p^k}$,
Appl. Math. Lett. {\bf 21} (2008), 1129--1133.

\bibitem{s3}  Y. Cao, Y. Cao, H. Q. Dinh, F-W. Fu, J. Gao, S. Sriboonchitta, Constacyclic codes of length $np^s$ over $\mathbb{F}_{p^m} + u\mathbb{F}_{p^m} $, Adv. Math. Commun. {\bf 12} (2018), 231--262.

\bibitem{s4} Y. Cao, On constacyclic codes over finite chain rings,
Finite Fields Appl. {\bf 24} (2013), 124--135.

\bibitem{s5} Y. Cao, Y. Cao, J. Gao, On a class of $(\delta+\alpha u^2)$-constacyclic codes over $\mathbb{F}_{q}[u]/\langle u^4\rangle$, IEICE Trans. Fundamentals, vol.E99-A, no.7 (2016), 1438--1445.

\bibitem{s6} Y. Cao, Y. Cao, L. Dong, Complete classification of $(\delta + \alpha u^2)$-constacyclic codes
over $\mathbb{F}_{3^m}[u]/\langle u^4\rangle$ of length $3n$,  Appl. Algebra in Engrg. Comm. Comput.
{\bf 29} (2018),
13--39.

\bibitem{s7} Y. Cao, Y. Cao, F. Ma, Complete classification of $(\delta + \alpha u^2)$-constacyclic codes over
$\mathbb{F}_{2^m}[u]/\langle u^4\rangle$ of length $2n$, Discrete Math. {\bf 340} (2017), 2840--2852.

\bibitem{s8} Y. Cao, Y. Gao, Repeate root cyclic $\mathbb{F}_q$-linear codes over
$\mathbb{F}_{q^l}$,
Finite Fields Appl. {\bf 31} (2015), 202--227.

\bibitem{s9} Y. Cao, Y. Cao, H. Q. Dinh, F-W. Fu, J. Gao, S. Sriboonchitta, A class of linear codes of length 2 over finite chain rings, J. Algebra Appl., April 2019, DOI: 10.1142/S0219498820501030

\bibitem{s10} Y. Cao, Y. Cao, H. Q. Dinh, F-W. Fu, Y. Gao, S. Sriboonchitta,
Type $2$ constacyclic codes over $\mathbb{F}_{2^m}[u]/\langle u^3\rangle$ of oddly even
length, Discrete Math. {\bf 342} (2019), 412--426.

\bibitem{s11} Y. Cao, Y. Cao, H. Q. Dinh, F-W. Fu, J. Gao, S. Sriboonchitta, A class of repeated-root constacyclic codes over $\mathbb{F}_{p^m}[u]/\langle u^e\rangle$ of Type $2$,
Finite Fields Appl. {\bf 55} (2019), 238--267.


\bibitem{s10} H. Q. Dinh, S. Dhompongsa, S. Sriboonchitta, Repeated-root constacyclic codes of prime power length over  $\frac{\mathbb{F}_{p^m}[u]}{\langle u^a\rangle}$
and their duals, Discrete Math. {\bf 339} (2016), 1706--1715.

\bibitem{s11} H. Q. Dinh, Constacyclic codes of length $2^s$ over
Galois extension rings of $\mathbb{F}_2+u\mathbb{F}_2$, IEEE Trans.
Inform. Theory {\bf 55} (2009), 1730--1740.

\bibitem{s12} H. Q. Dinh, Constacyclic codes of length $p^s$ over
$\mathbb{F}_{p^m}+u \mathbb{F}_{p^m}$, J. Algebra {\bf 324} (2010),
940--950.

\bibitem{s13} H. Q. Dinh, L. Wang, S. Zhu, Negacyclic codes of length $2p^s$ over $\mathbb{F}_{p^m}+u \mathbb{F}_{p^m}$, Finite Fields Appl., {\bf 31} (2015), 178-201.

\bibitem{s14} H. Q. Dinh, S. Dhompongsa, and S. Sriboonchitta, On constacyclic codes of length
$4p^s$ over $\mathbb{F}_{p^m}+u \mathbb{F}_{p^m}$, Discrete Math. {\bf 340} (2017), 832--849.

\bibitem{s15} H. Q. Dinh, A. Sharma, S. Rani, and S. Sriboonchitta, Cyclic and negacyclic codes of length
$4p^s$ over $\mathbb{F}_{p^m}+u \mathbb{F}_{p^m}$, J. Algebra Appl., DOI 10.1142 /S0219498818501736, January 2018.

\bibitem{s16} H. Q. Dinh, S. R. L\'{o}pez-Permouth, Cyclic and negacyclic codes over finite chain rings,
IEEE Trans. Inform. Theory {\bf 50} (2004), 1728--1744.

\bibitem{s17} S. T. Dougherty, P. Gaborit, M. Harada, P. Sole, Type II codes over $\mathbb{F}_2+u\mathbb{F}_2$,
IEEE Trans. Inform. Theory {\bf 45} (1999), 32--45.

\bibitem{s18}  S. T. Dougherty,  J-L. Kim,  H. Kulosman,  H. Liu: Self-dual
codes over commutative Frobenius rings, Finite Fields Appl. {\bf 16} (2010), 14--26.

\bibitem{s19} W. C. Huffman, On the decompostion of self-dual codes over $\mathbb{F}_2+u\mathbb{F}_2$
with an automorphism of odd prime number, Finite Fields Appl. {\bf 13} (2007), 682--712.

\bibitem{s20} X. Kai, S. Zhu, P. Li, $(1+\lambda u)$-constacyclic codes over $\mathbb{F}_p[u]/\langle u^k\rangle$,
J. Franklin Inst. {\bf 347} (2010), 751--762.

\bibitem{s23} H. R. Mahmoodi, R. Sobhani, On some constacyclic codes over the ring $\mathbb{F}_{p^m}[u]/\langle u^4\rangle$,
Discrete Math. {\bf 341} (2018), 3016--3122.

\bibitem{s21} G. Norton,  A. S\u{a}l\u{a}gean-Mandache, On the structure of linear and cyclic
codes over finite chain rings, Appl. Algebra in Engrg. Comm. Comput.
{\bf 10} (2000), 489--506.

\bibitem{s22} J. F. Qian, L. N. Zhang, S. Zhu, $(1+u)$-constacyclic and cyclic codes over $\mathbb{F}_2+u\mathbb{F}_2$,
Appl. Math. Lett. {\bf 19} (2006), 820--823.

\bibitem{s23} R. Sobhani, Complete classification of $(\delta+\alpha u^2)$-constacyclic
codes of length $p^k$ over $\mathbb{F}_{p^m}+u\mathbb{F}_{p^m}+u^2\mathbb{F}_{p^m}$,
Finite Fields Appl. {\bf 34} (2015), 123--138.

\bibitem{s24} W. Zhao, X. Tang, Z. Gu, All $\alpha+ u\beta$-constacyclic codes of length $np^s$ over
$\mathbb{F}_{p^m}+u\mathbb{F}_{p^m}$, Finite Fields Appl. {\bf 50} (2018), 1--16.

\end{thebibliography}




\end{document}